\documentclass{article}
\usepackage{graphicx}
\usepackage{amsmath,amssymb}
\usepackage{graphics,color,array,calc,rotating,epsfig,psfrag}
\numberwithin{equation}{section}
\usepackage{subcaption}
\usepackage{cite}
\usepackage{bm}
\usepackage{dcolumn}
\usepackage{float}
\usepackage{amsfonts}
\usepackage[mathscr]{eucal}
\def\be{\begin{equation}} \def\ee{\end{equation}}
\def\bea{\begin{eqnarray}} \def\eea{\end{eqnarray}}

\newcommand\prt{\partial}

\newcommand{\nn}{\nonumber}
\begin{document}
\baselineskip 18pt%
\begin{titlepage}
\vspace*{1mm}%
\hfill%
\vspace*{15mm}%
\hfill
\vbox{
    \halign{#\hfil         \cr
%           hep-th/yymmnnn\cr
%         IPM/P-2009/025  \cr
          } % end of \halign
      }  % end of \vbox
\vspace*{20mm}

\begin{center}
{\large {\bf  Propagation of a scalar field with non-minimal coupling in three dimensions: Hawking radiation and Quasi normal modes}}\\
\vspace*{5mm}
{Davood Mahdavian Yekta \footnote{d.mahdavian@hsu.ac.ir}, Morteza Shariat}\\
\vspace*{0.2cm}
{ Department of Physics, Hakim Sabzevari University, P.O. Box 397, Sabzevar, Iran}\\
\vspace*{1cm}
\end{center}

\begin{abstract}

In this paper we investigate an exact spectrum of quasi normal modes (QNMs) for perturbations of a scalar field  coupled non-minimally with the Einstein tensor of an uncharged, non-rotating Banados, Teitelboim, and Zanelli (BTZ) black hole in three-dimensional spacetime. Due to the geometry around the black hole, the scalar field encounters an effective potential barrier. We study this potential numerically and derive exact numerical results for the greybody factors (GFs) and discuss their profiles in terms of the coupling constant and black hole parameters.  We then proceed to derive the Hawking radiation spectrum for BTZ black hole.

\end{abstract}
\end{titlepage}

%%%%%%%%%%%%%%%%%%%%%%%%%%%%%%%%%%%
\section{Introduction}
Recently it has been devoted a lot of studies of gravitational theories to modification of Einstein's gravity. One class of these theories concern the scalar-tensor theories, such as Horndeski theory which gives a second order field equations in four dimensions\cite{Horndeski:1974wa,Nicolis:2008in,Deffayet:2009wt}. The Lagrangian of this model contains a term for coupling of a scalar field with curvature tensors. This kind of coupling has interesting cosmological implications\cite{Amendola:1993uh, Capozziello:1999xt,Sushkov:2009hk, Saridakis:2010mf}. The coupling of the scalar field to Einstein tensor can be regarded effectively as a cosmological constant \cite{Sushkov:2009hk}. On the other hand, this scalar-tensor concept is also intriguing in three-dimensional general relativity (GR). Vanishing Newtonian potential in three-dimensional GR disqualifies it to serve as a compatible theory for three-dimensional gravity while the proper description is obtained by dimensional reduction of four-dimensional GR to a scalar-tensor theory in three dimensions. The other standard approach which we will not discuss it here is the addition of higher derivative corrections to GR to produce massive gravity theories in three dimensions proposed in Refs. \cite{Deser:1981wh,Deser:1982vy,Bergshoeff:2009hq,Bergshoeff:2014pca,Gullu:2010pc}.  The nice feature of these theories is that there are different kinds of black hole solutions to their equations of motion \cite{Moussa:2003fc}-\!\!\cite{Ahmedov:2010uk}, however the most reputed asymptotically AdS one is the BTZ black hole \cite{Banados:1992wn}. Within the framework of this theory, the existence of black holes is anticipated, which is an appropriate ground for understanding many aspects of gravity theory.

The stability of a black hole can be examined by the study of dynamical behaviors of the perturbations in its background spacetime. The natural vibrational modes of these perturbations in the spacetime exterior to an event horizon are called quasi normal modes (QNMs). The corresponding frequency spectrum of the QNMs is discrete and complex. The imaginary part of the frequency signals the presence of damping, a necessary consequence of boundary conditions that require energy to be carried away from the system. QNMs are powerful tools for studying the evolution of perturbations outside the black holes \cite{Regge:1957td,Zerilli:1970se,Teukolsky:1972my,Ferrari:1984zz,Aros:2002te} and the AdS/CFT interpretations in the semi-classical considerations \cite{Chan:1996yk,Das:1999pt,Birmingham:2001hc,Birmingham:2001pj,Son:2002sd,Sachs:2008gt}. The quasi normal frequencies correspond to the poles of
the retarded correlation function in the dual conformal field theory and their imaginary part determines the relaxation time scale back to the thermal
equilibrium. The QNMs can also be used to identify the presence of  black holes. Recently that the existence of gravitational waves has been detected by LIGO \cite{Abbott:2016blz,Abbott:2016nmj}, the existence of black holes and their integration have been proven, and QNMs are more and more of a concern (for typical reviews see Refs. \cite{Nollert:1999ji,Kokkotas:1999bd,Berti:2009kk,Konoplya:2011qq}). QNMs are determined in terms of black hole parameters and constants of the theory.

In the context of quantum field theory for black hole backgrounds, Hawking showed that they can emit radiation with a characteristic like a black body spectrum, known as black hole Hawking radiation \cite{Hawking:1974sw,Hawking:1974rv,Hawking:1982dh}. Though the spectrum of a black hole radiation at the event horizon is exactly that of a black body, but when quantum mechanics is taken into account, black holes are in fact not black. It means that the initial radiation will get modified by the non-trivial spacetime geometry that the black hole generates when it travels away from event horizon. In fact, for an observer located at infinity this spectrum differs from that of a black body by a coefficient called greybody factor (GF), or else absorption cross section, a frequency- and geometry-dependent quantity that filters the initial Hawking radiation \cite{Das:1996we,Brady:1996za,Gubser:1997yh,Kanti:2014dxa}.
QNMs can also be considered as the poles in the black hole GFs which are important in the study of Hawking radiation\cite{Motl:2003cd}. Studying GFs in different frequency regimes enable us to learn more about the quantum structure of black holes and consequently the quantum gravity. In the context of AdS/CFT \cite{Maldacena:1997re,Gubser:1998bc,Witten:1998qj,Aharony:1999ti}, the decay of a test field in the black hole spacetime describes the return of a perturbed thermal state to its thermal equilibrium in the CFT \cite{Horowitz:1999jd,Konoplya:2002zu,Cardoso:2001bb}.

In the present work we study possible instabilities of a non-rotating BTZ black hole by calculating the QNMs of perturbations of a scalar field coupled non-minimally to Einstein tensor. The coupling constant emerges in the effective potential which means that coupling between the scalar perturbation and Einstein tensor will change the dynamical evolution of the scalar perturbation in the background spacetime. We investigate the GFs by calculating the transmission and reflection coefficients when the scalar perturbations encounter the potential barrier outside the event horizon of BTZ black hole. To our knowledge the GFs of BTZ black holes has been calculated for a scalar field non-minimally coupled with Ricciscalar tensor in Ref. \cite{Panotopoulos:2016wuu} not Einstein tensor, and they considered a little detail of stability. On the other hand, our results can be useful to compare with the perturbations around the black holes which are not asymptotically AdS, such as warped AdS \cite{Moussa:2003fc,Anninos:2008fx} or Lifshitz geometries \cite{AyonBeato:2009nh}.

There is a lot of studies about the perturbations of physical fields, which are minimally or non-minimally coupled to the curvature or Maxwell tensors, in the vicinity of different kinds of black holes in a range of gravitational theories. For example the study of QNMs and GFs for scalar perturbations in the minimal coupling are given in Refs. \cite{Maldacena:1996ix}-\!\!\cite{Yu:2018zqd} and in the case of non-minimal coupling in Refs. \cite{Chen:2010ru}-\!\!\cite{Minamitsuji:2014hha}. There are also some works in three-dimensional theory for minimal \cite{Cardoso:2001hn}-\!\!\cite{Rincon:2018sgd} and non-minimal \cite{Yao:2011kf,Panotopoulos:2018can} coupling of a scalar with Ricciscalar curvature around the BTZ and Warped AdS black holes.

The layout of the rest of this paper is as follows; In Sec. 2, we briefly introduce the non-minimal coupling theory of a scalar field to the Einstein tensor in three dimensions. In Sec. 3, we study the scalar perturbations by solving the field equation in a non-rotating BTZ background analytically. We obtain an expression for the effective potential near the black hole and plot it as a function of different parameters. We also determine exact spectrum for QNM frequencies of the scalar field scattered from the black hole and give a comment on the AdS/CFT interpretation of these modes. In Sec. 4, we calculate the GFs and provide a discussion about the decay rate of black holes known as Hawking radiation. The last section is devoted to giving the brief summary and conclusion.
%%%%%%%%%%%%%%%%%%%%%%%%%%%%%%%%%%%
%@@@@@@@@@@@@@@@@@@@@@@@@@@@@@@@@@@@@
\section{Non-minimal coupling of scalar field to Einstein tensor}

This kind of coupling between a scalar field and linear curvature tensors of gravity have been proposed extensively in the cosmological models\cite{Amendola:1993uh,Capozziello:1999xt,Sushkov:2009hk}. The most general Lagrangian of gravity which is linear in $\phi$ and curvature tensors introduced in \cite{Amendola:1993uh}, includes the following terms
\be \label{NML} L\!=\!\alpha_1 \prt^{\mu} \phi \prt_{\mu} \phi\, R+\alpha_2 \prt^{\mu} \phi \prt^{\nu} \phi\, R_{\mu\nu}+\alpha_3 \phi \Box \phi\, R+\alpha_4 \phi \nabla^{\mu} \nabla^{\nu} \phi\, R_{\mu\nu}+\alpha_5  \phi \prt^{\mu} \phi \prt_{\mu} R+\alpha_6 \phi^2 \Box R\,.\ee
 In \cite{Amendola:1993uh}, it has been shown that the last three terms can be neglected due to some Bianchi identities and divergencies, and the third term can be removed because of boundary conditions, so only the first two terms remain to consider. For a particular choice of couplings, $\alpha_1=-\frac12 \alpha_2$, we obtain a non-minimal coupling of scalar field to Einstein tensor. Though this Lagrangian has astronomical and cosmological implications in four dimensions, but it is of great interest for us to consider this kind of coupling in three dimensions which might has an important role in our understanding of quantum theory of gravity.

In this paper, we consider a theory where the first order derivative of the
scalar field is non-minimally coupled to the Einstein tensor and its action is described by
\be \label{sact} S=\int d^3 x\,\sqrt{-g} \Big[-\frac12 (\prt_{\mu} \phi)^2-\frac12 m_s^2 \phi^2+\frac{\xi}{2} G^{\mu\nu} \prt_{\mu} \phi \prt_{\nu} \phi \Big]\,, \ee
which in comparison to relation (\ref{NML}) we have chosen the couplings as
\be\label{coup}\alpha_1=-\frac12 \alpha_2=-\frac{\xi}{4}, \ee
where $\xi$ is a coupling constant with the dimension of length-squared, $m_s$ is the mass of scalar field, and $G^{\mu\nu}$  is the inverse of Einstein tensor. This coupling is essentially different from the non-minimal coupling of scalar field with Ricciscalar tensor in \cite{Panotopoulos:2016wuu}, in particular when we consider the metrics which are not asymptotically AdS. Varying the action (\ref{sact}) with respect to $\phi$, we obtain the scalar field equation of motion
\be\label{eom} \frac{1}{\sqrt{-g}} \prt_{\mu}\Big[\sqrt{-g} (g^{\mu\nu}-\xi G^{\mu\nu})\prt_{\nu}\Big]\phi-m_s^2 \phi =0,\ee
which is a modified Klein-Gordon equation in comparison to general quantum field theory and $g^{\mu\nu}$ is the inverse of spacetime metric. It must be noticed that though the scalar field equation of (\ref{sact}) is higher order of derivatives, but as discussed in \cite{Sushkov:2009hk}, by using the Bianchi identity $\nabla^{\mu} G_{\mu\nu}=0$ and the parameter selection (\ref{coup}) one can show that the equation is of second order, just as Horndeski Lagrangian \cite{Horndeski:1974wa} which leads to a second order equation of motion.

It is a hard work in general relativity to obtain an exact analytic solution for the equations of motion of the action (\ref{sact}), since we should consider the back-reaction effect of the dynamical scalar field matter on the geometry of space time background. However, we assume its back-reaction effect on the background can be neglected exactly and we can study the perturbations of this scalar field around a black hole off arbitrary geometry. In fact the study of black holes physics, as fascinating objects in modern physics, is of interest and exploring the behavior of scalar fields, when are non-minimally coupled to Einstein tensor in their backgrounds, is constructive. Here, we will use the equation (\ref{eom}) to investigate the QNMs, GFs and Hawking radiation for a scalar field perturbations in the background of BTZ black holes by using the matching technique, which has been widely used in literatures as in Refs. \cite{Das:1996we,Yao:2011kf,Pappas:2016ovo,Chen:2010ru}.

%@@@@@@@@@@@@@@@@@@@@@
\section{The effective potential of non-rotating BTZ black Hole}
The uncharged, non-rotating BTZ black hole that we consider is a solution of the Einstein
gravity coupled to a cosmological constant $\Lambda=-\frac{1}{l^2}$ in three dimensions. The line element of the BTZ black hole is given by
\be \label{BTZ} ds^2=-f(r) dt^2+f(r)^{-1} dr^2+r^2 (d\varphi+N^{\varphi}(r) dt)^2\,,\ee
where the functions $f(r)$ and $N^{\varphi}(r)$ is defined in terms of the ADM mass and angular momentum of BTZ black hole as
\be f(r)=-M+\frac{r^2}{l^2}+\frac{J^2}{4r^2},\qquad N^{\varphi}(r)=\frac{J}{2r^2},\nn\ee
where in the non-rotating limit $J\rightarrow 0$, we have $f(r)=-M+\frac{r^2}{l^2}$ and $N^{\varphi}(r)=0$. The BTZ black holes have asymptotically locally AdS$_{3}$ isometry and the parameter $l$ defines the size of AdS space. The Hawking temperature of the black hole \cite{Bardeen:1973gs} is given by
\be\label{HT} T_H=\frac{\kappa}{2\pi}=\frac{1}{4\pi} f'(r)\Big|_{r=r_h}=\frac{\sqrt{M}}{2\pi l}\,,\ee
where  $r_h=l \sqrt{M}$ is the event horizon of the black hole with area $A_h=2\pi r_h$ and $\kappa$ is the surface gravity on the horizon. We use this value in Sec. 4 to determine the rate of BTZ black hole decay.

Now we treat the weak external scalar filed as a probe field, and then study the effects of the coupling constant $\xi$ on the quasi normal frequencies of perturbations of this scalar filed in the background of a non-rotating BTZ black hole. Substituting the following factorized ansatz for the massive scalar field
 \be \label{ansatz}\phi(t,r,\varphi)=e^{-i \omega t+i m \varphi} R(r),\ee
and the metric (\ref{BTZ}) for $J=0$ in the equation (\ref{eom}), we obtain a second order differential equation for the radial part as
\be\label{req} R''+\left(\frac{1}{r}+\frac{f'}{f}\right)R'+\left(\frac{\omega^2}{f^2}-\frac{m^2}{r^2 f}-\frac{2r\, m_s^2}{(2r-\xi f') f}\right)R=0, \ee
where the prime is the derivative with respect to the r and $m$ is a quantum number of the angular coordinate $\varphi$, and $\omega$ is the frequency of scalar field perturbation. We can write the equation (\ref{req}) as a Schr$\ddot o $dinger-like equation \cite{Regge:1957td,Zerilli:1970se} with an effective potential. To finding this potential we need to introduce new definitions by
\be \label{tc} R(r)=r^{\frac{2-d}{2}} \psi(r)\,,\qquad x=\int \frac{dr}{f(r)}\,,\ee
for a d-dimensional spacetime. The new radial coordinate $x$ is called the tortoise coordinate which its integration from event horizon to some distance $r$ gives
\be x=\frac{l^2}{2r_{h}} \ln\left(\frac{r-r_h}{r+r_h}\right).\ee
After some calculation and substituting (\ref{tc}) in the equation (\ref{req}) we find that
\be \label{seq} \frac{d^2 \psi}{dx^2}+(\omega^2-V_{eff})\psi=0\,,\ee
where the effective potential is
\be\label{eff} V_{eff}(r)=f(r)\left(\frac{4r^2 f'-2r\xi f'^2+\xi f f'-2r f+8r(m^2+r^2 m_s^2)-4m^2 \xi f'}{4r^2(2r-\xi f')}\right).\ee
As is obvious this effective potential depends on the constant parameters of the theory $m,m_s,\xi$, and the black hole parameters $M,l$. Here, we discuss and analyze this potential numerically. The potential can have some local maximum regarding to different values of constants. Extremizing the potential (\ref{eff}) with respect to $r$ and then substituting $f(r)=-M+r^2/l^2$, we obtain the location of the extremum as
\be\label{ex} r=\Big[\frac{l^4 M (M+4m^2)  (\xi-l^2)}{4m_s^2 l^4-3(\xi-l^2)}\Big]^{\frac14},\ee
in which the scalar field has some instability. In order to have a real value for $r$ we must have the following conditions, simultaneously,
\be \label{excs} \xi> l^2 ,\qquad (4m_s^2 l^4+3l^2)>3\xi, \ee
otherwise there is no local extremum for the effective potential and hence the spacetime is stable. On the other words, we have an instability independent of the fact that the location of the maximum be inside or outside the horizon. Also (\ref{excs}) shows that in the case of massless scalar field there is no maximum and the black hole is stable\cite{Cardoso:2001hn,Govindarajan:2000vq}. The interest to finding a local maximum for the effective potential is an essential condition for our work, since we want to study the GFs of the black hole which it needs a potential barrier. In the set of figures \{\ref{f1},\ref{f2}\}, we analyze the behavior of effective potential for different numerical values of parameters;
%@@@@@@@@@@@@@@@@@
\begin{itemize}
\item The Figs. (\ref{fig1})-(\ref{fig6}) show that the presence of local maximum depends on the strength of non-minimally coupling constant $\xi$. As depicted in Fig. (\ref{fig1}), in the weak coupling limit, by increasing the value of AdS radius, we still have local maximums while for large couplings by increasing $l$, as shown in Fig. (\ref{fig6}), they were disappeared. We have not plotted the diagrams from a particular event horizon point, since for each value of $l$ there is an event horizon. But it is clear that for smaller (bigger) values of $l$ ($\xi$) the black hole is more stable than bigger (smaller) ones. That is the solid line is more stable than the dash line and so on.

\begin{figure}[H]
  \centering
  \begin{subfigure}{0.3\textwidth}
    \includegraphics[width=\textwidth]{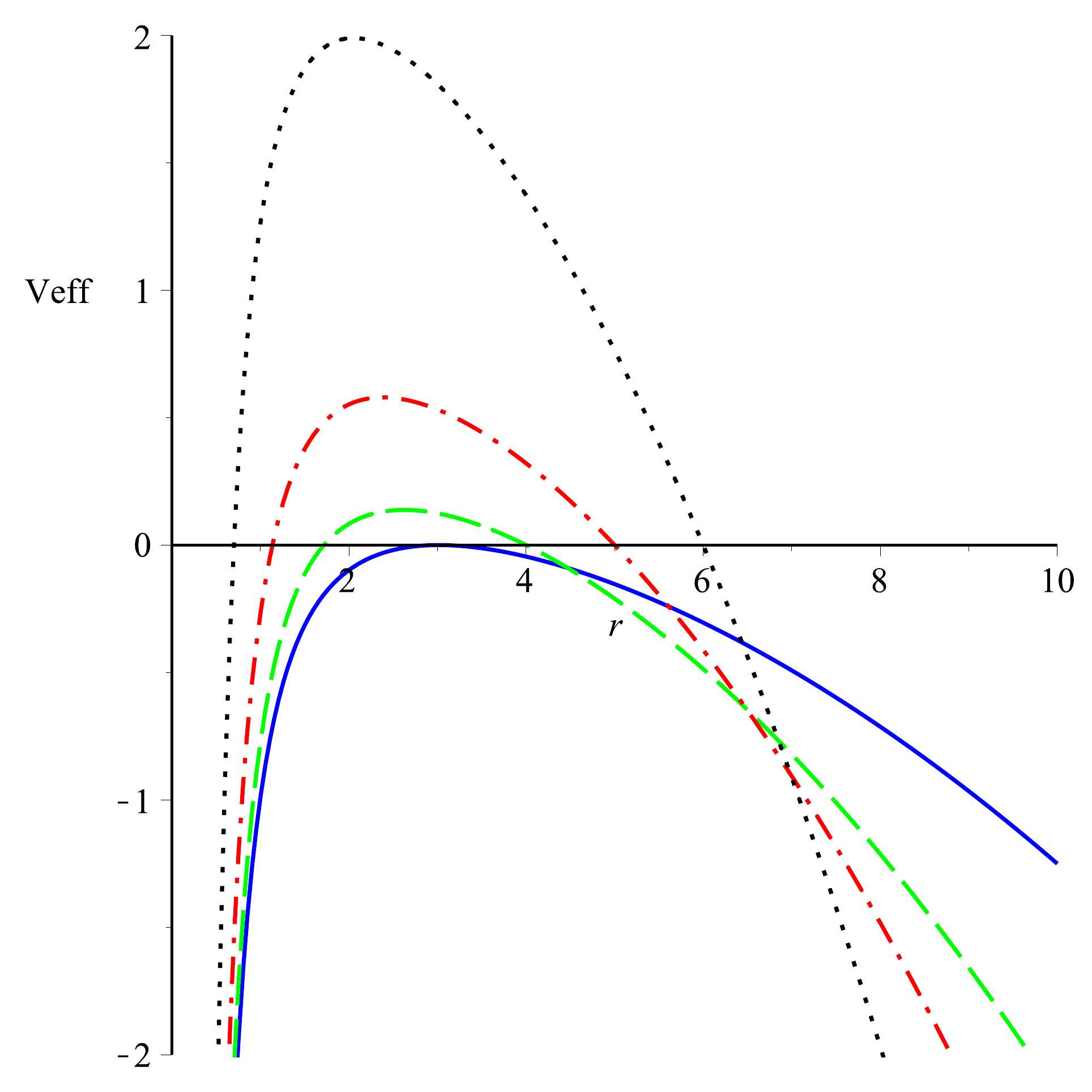}
    \caption{{\scriptsize $\xi\!=\!50$ and $l\!=\!3$(solid), $l\!=\!4$(dash), $l\!=\!5$(dashdot), $l\!=\!6$(dot)}}
    \label{fig1}
  \end{subfigure}
  \hspace{5mm}
  \begin{subfigure}{0.3\textwidth}
    \includegraphics[width=\textwidth]{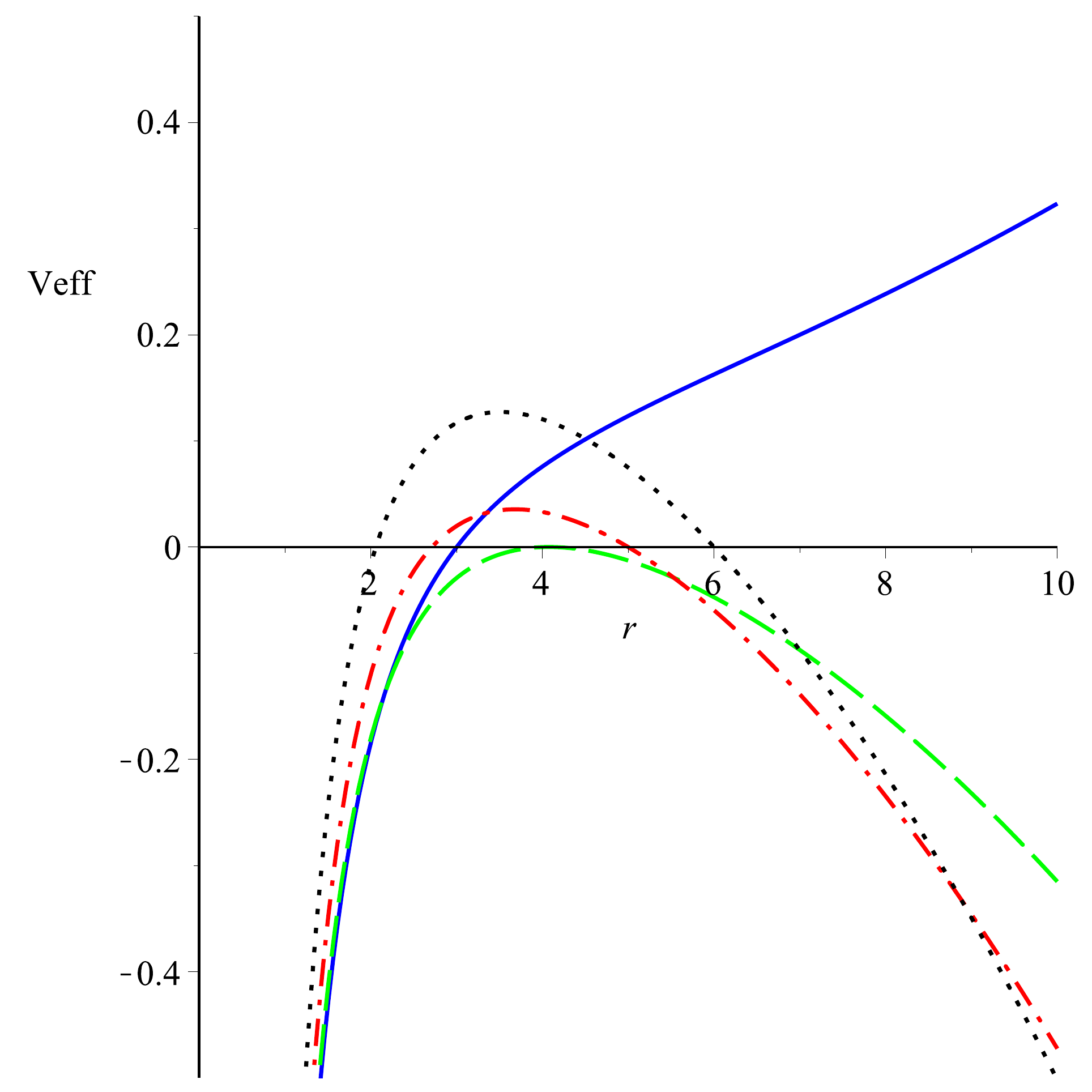}
    \caption{{\scriptsize $\xi\!=\!150$ and $l\!=\!3$(solid), $l\!=\!4$(dash), $l\!=\!5$(dashdot), $l\!=\!6$(dot)}}
    \label{fig2}
  \end{subfigure}
  \hspace{5mm}
   \begin{subfigure}{0.3\textwidth}
    \includegraphics[width=\textwidth]{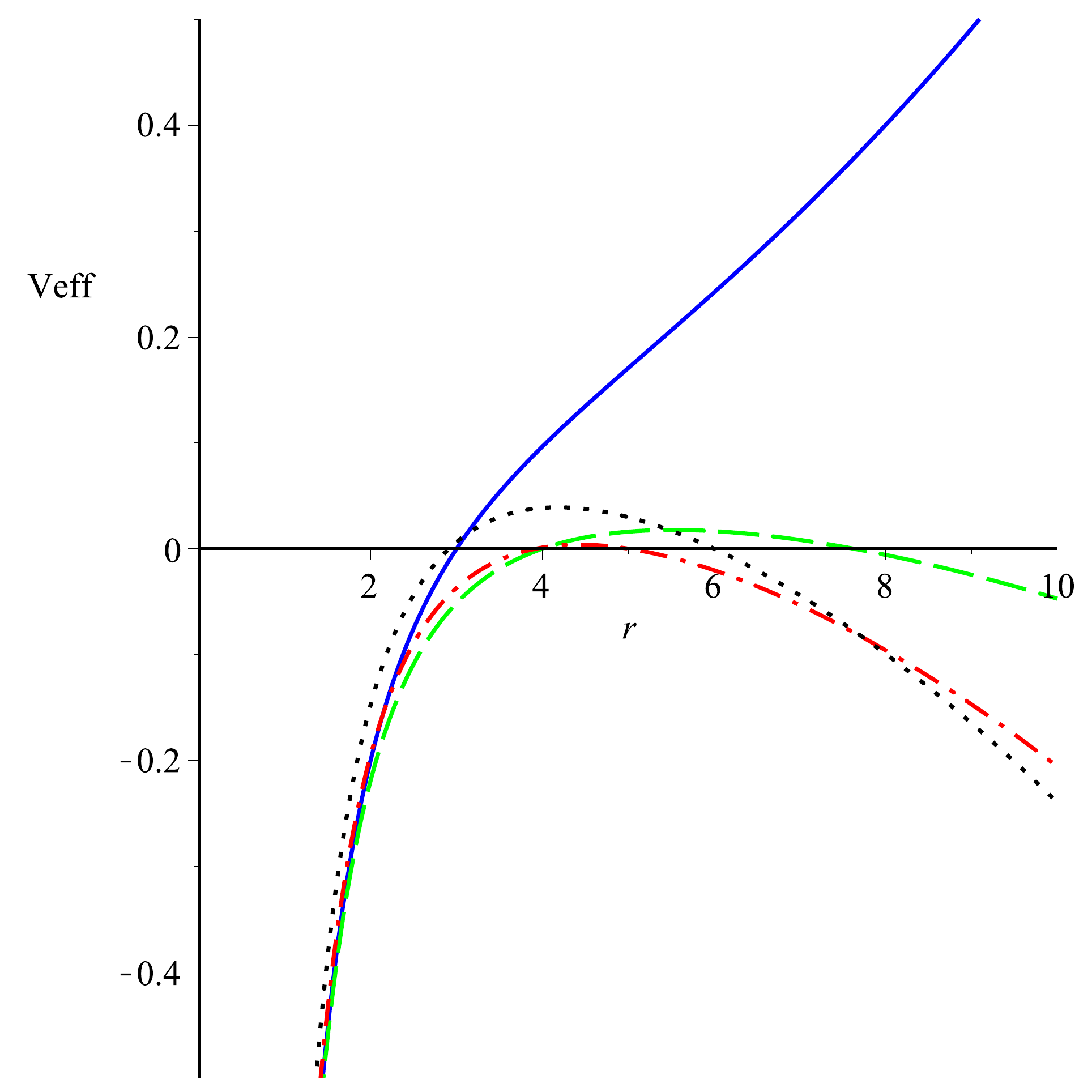}
    \caption{{\scriptsize $\xi\!=\!250$ and $l\!=\!3$(solid), $l\!=\!4$(dash), $l\!=\!5$(dashdot), $l\!=\!6$(dot)}}
    \label{fig3}
  \end{subfigure}
   \begin{subfigure}{0.3\textwidth}
    \includegraphics[width=\textwidth]{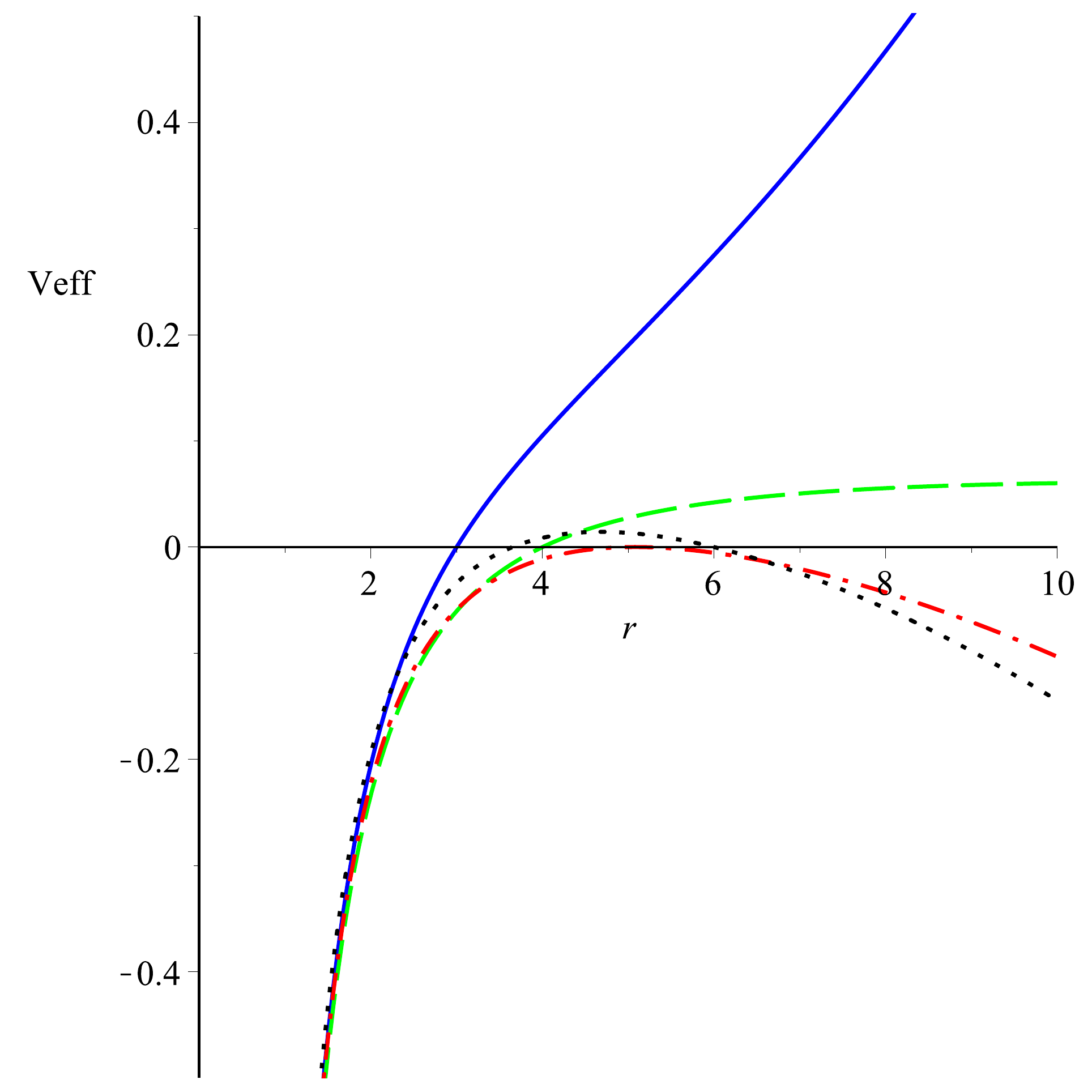}
    \caption{{\scriptsize $\xi\!=\!350$ and $l\!=\!3$(solid), $l\!=\!4$(dash), $l\!=\!5$(dashdot), $l\!=\!6$(dot)}}
    \label{fig4}
  \end{subfigure}
  \hspace{5mm}
   \begin{subfigure}{0.3\textwidth}
    \includegraphics[width=\textwidth]{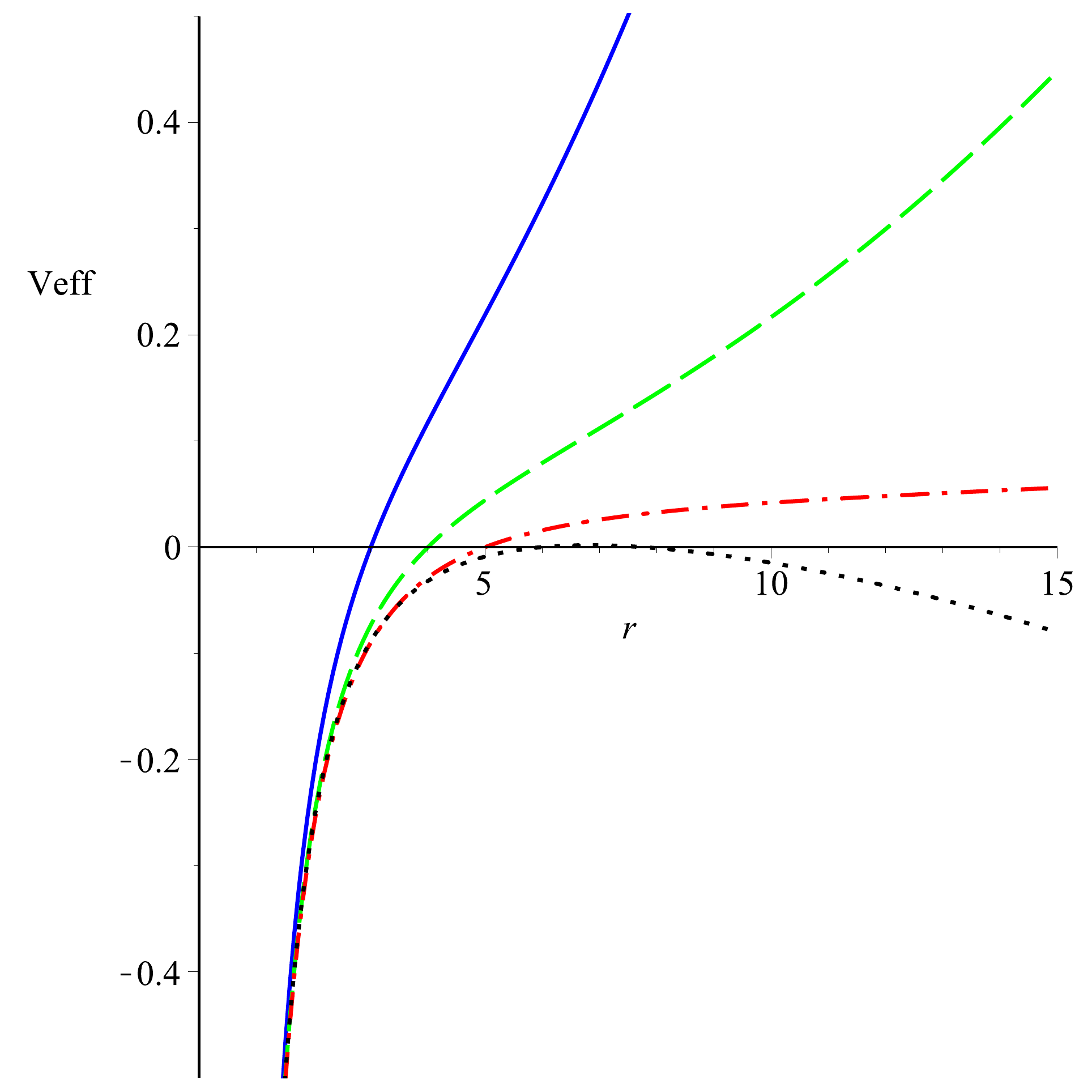}
    \caption{{\scriptsize $\xi\!=\!900$ and $l\!=\!3$(solid), $l\!=\!4$(dash), $l\!=\!5$(dashdot), $l\!=\!6$(dot)}}
    \label{fig5}
  \end{subfigure}
  \hspace{5mm}
   \begin{subfigure}{0.3\textwidth}
    \includegraphics[width=\textwidth]{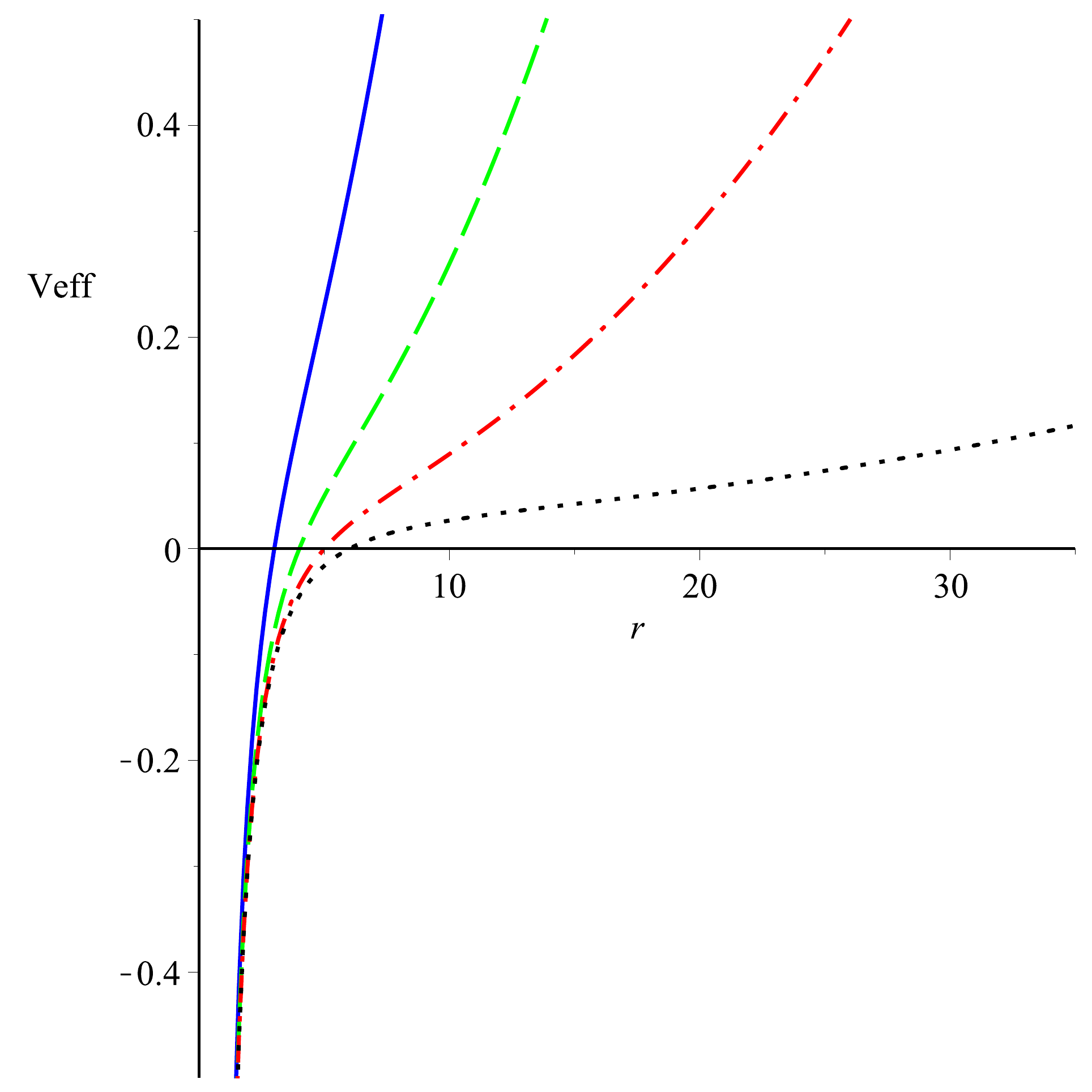}
    \caption{{\scriptsize $\xi\!\!=\!\!2000$ and $l\!=\!3$(solid), $l\!=\!4$(dash), $l\!=\!5$(dashdot), $l\!=\!6$(dot)}}
    \label{fig6}
  \end{subfigure}
  \caption{The effective potential for different coupling limits when $m=M=m_s=1$ and the location of event horizon is at $r_{+}=l$.}
  \label{f1}
\end{figure}
%@@@@@@@@@@@@@@@@@@@@@@@@@@@
\item In the second kind of diagrams in Figs.(\ref{fig7})-(\ref{fig10}), we have chosen $l=5$. In brief, in Fig.(\ref{fig7}) we consider different values of $m_s$, in Fig.(\ref{fig8}) for black hole masses $M$, in Fig.(\ref{fig9}) for some quantum numbers $m$, and in Fig.(\ref{fig10}) versus different couplings $\xi$. Regarding to these plots we conclude that;

\item[-] It can be inferred from Fig.(\ref{fig7}) that the mass of scalar field has a determinative rule in the stability of the black hole. That is, by increasing the mass the slope of the potential becomes more negative and the system goes to unstable phase.

\item[-] According to the Figs.(\ref{fig7}),  (\ref{fig9}), and  (\ref{fig10}) for any arbitrary value of black hole mass $M$, there is an intersection point among different plots of some particular parameter at distance $r=5$. In fact this is the event horizon for which the effective potential vanishes. According to Fig.(\ref{fig8}), though the mass of the black hole has no effect on the existence of local maximum, but its value scales the location of extremum and the magnitude of effective potential.

\begin{figure}[H]
\centering
  \begin{subfigure}{0.4\textwidth}
    \includegraphics[width=\textwidth]{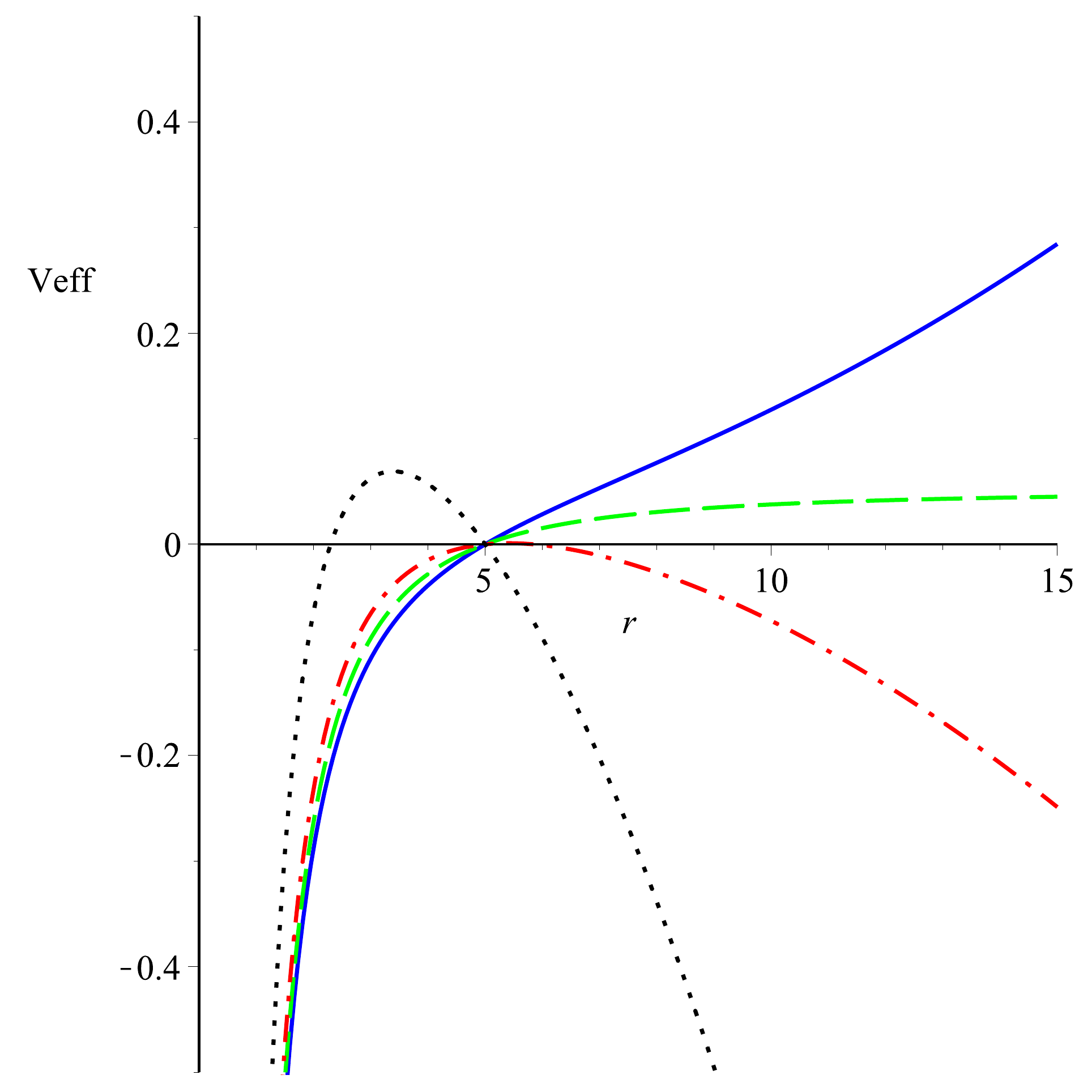}
    \caption{{\scriptsize $m\!=\!1$, $M\!=\!1$, $\xi\!=\!400$, $l\!=\!5$, $m_s\!=\!0$(solid),\\ $m_s\!=\!0.65$(dash), $m_s\!=\!1$(dashdot), $m_s\!=\!2$(dot)}}
    \label{fig7}
  \end{subfigure}
  \hspace{5mm}
  \begin{subfigure}{0.4\textwidth}
    \includegraphics[width=\textwidth]{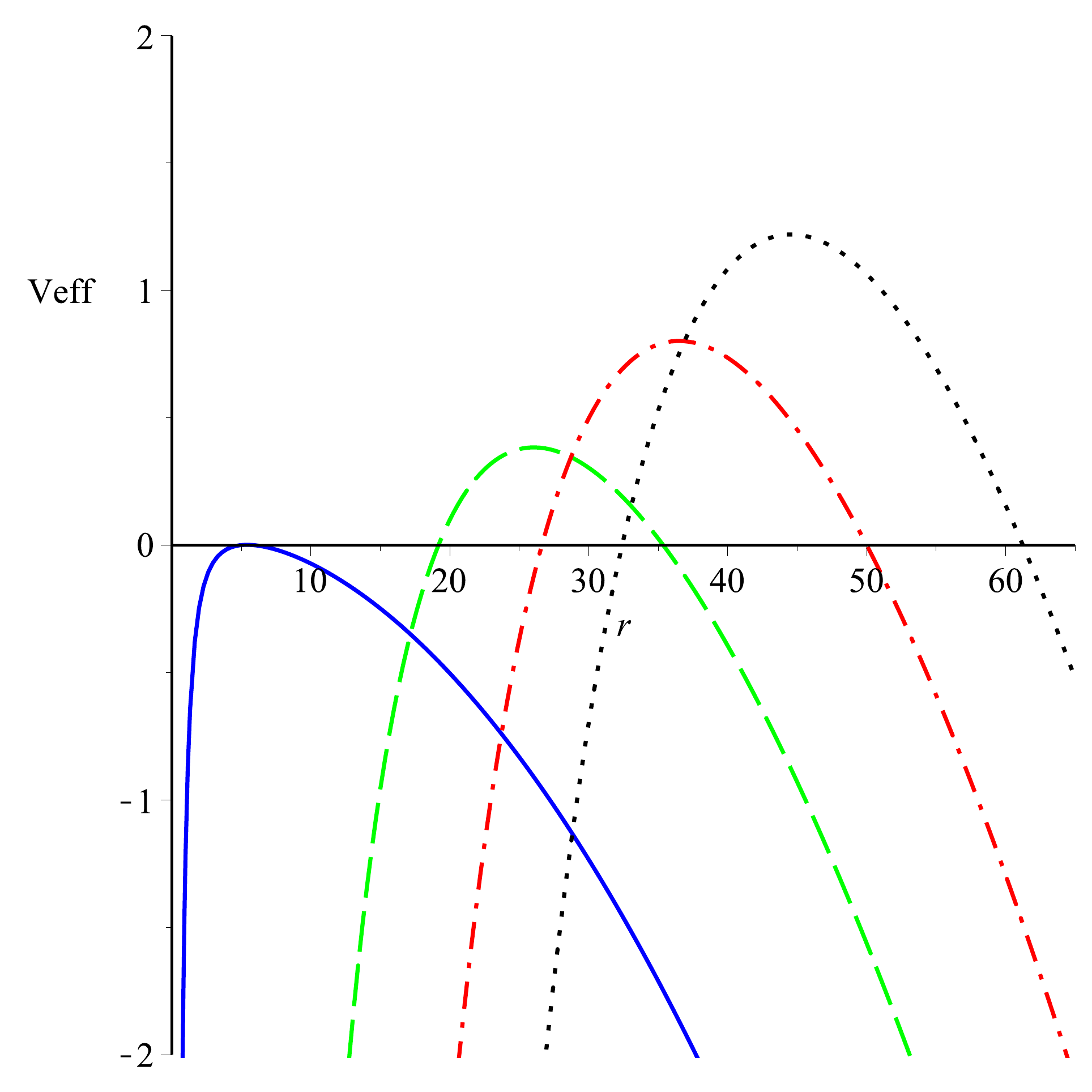}
    \caption{{\scriptsize $m\!=\!1$, $\xi\!=\!400$, $m_s\!=\!1$, $l\!=\!5$, $M\!=\!1$(solid),\\ $M\!=\!50$(dash), $M\!=\!100$(dashdot), $M\!=\!150$(dot)}}
    \label{fig8}
  \end{subfigure}
   \begin{subfigure}{0.4\textwidth}
    \includegraphics[width=\textwidth]{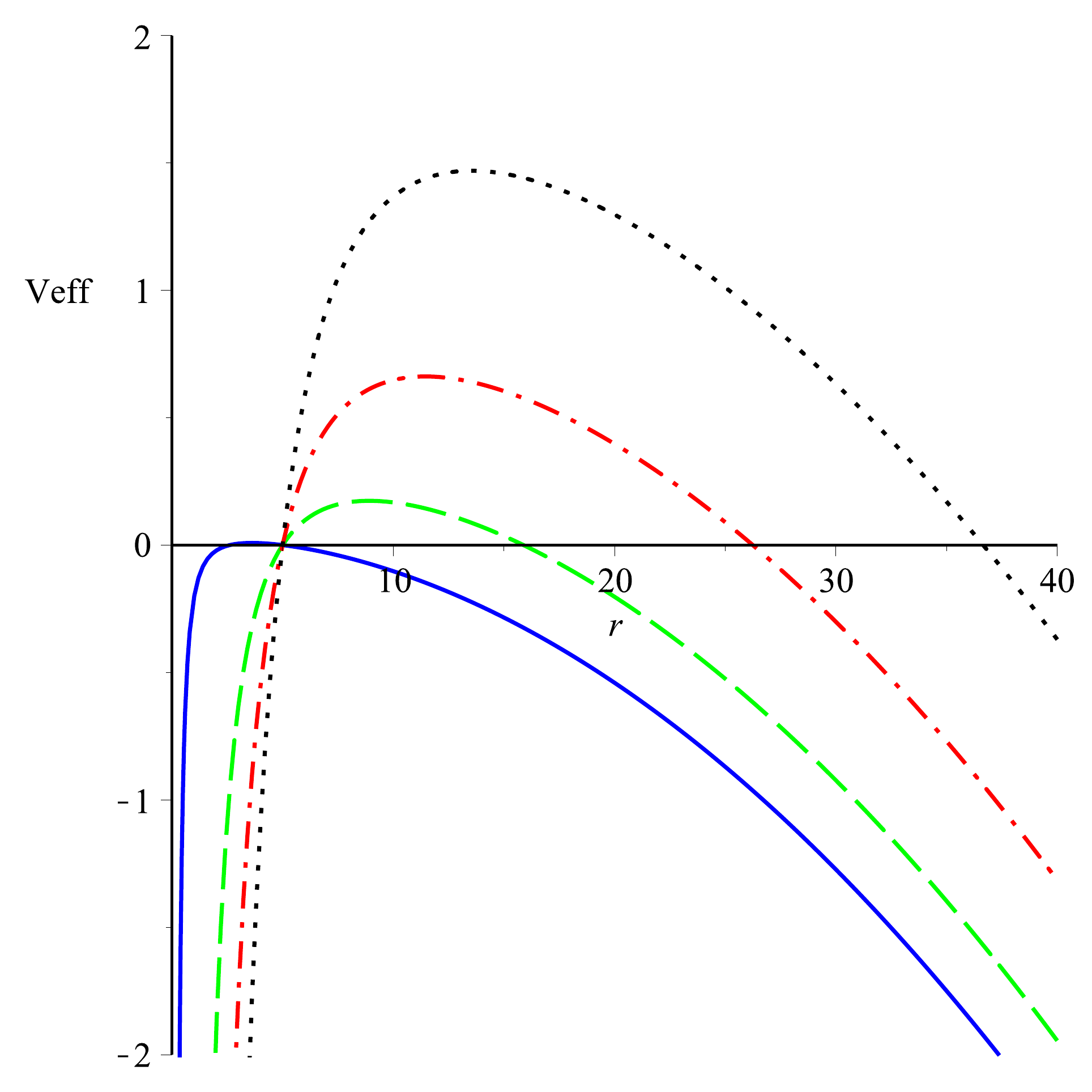}
    \caption{{\scriptsize $\xi\!=\!400$, $M\!=\!1$, $m_s\!=\!1$, $l\!=\!5$, $m\!=\!0$(solid),\\ $m\!=\!3$(dash), $m\!=\!5$(dashdot), $m\!=\!7$(dot)}}
    \label{fig9}
  \end{subfigure}
  \hspace{5mm}
    \begin{subfigure}{0.4\textwidth}
    \includegraphics[width=\textwidth]{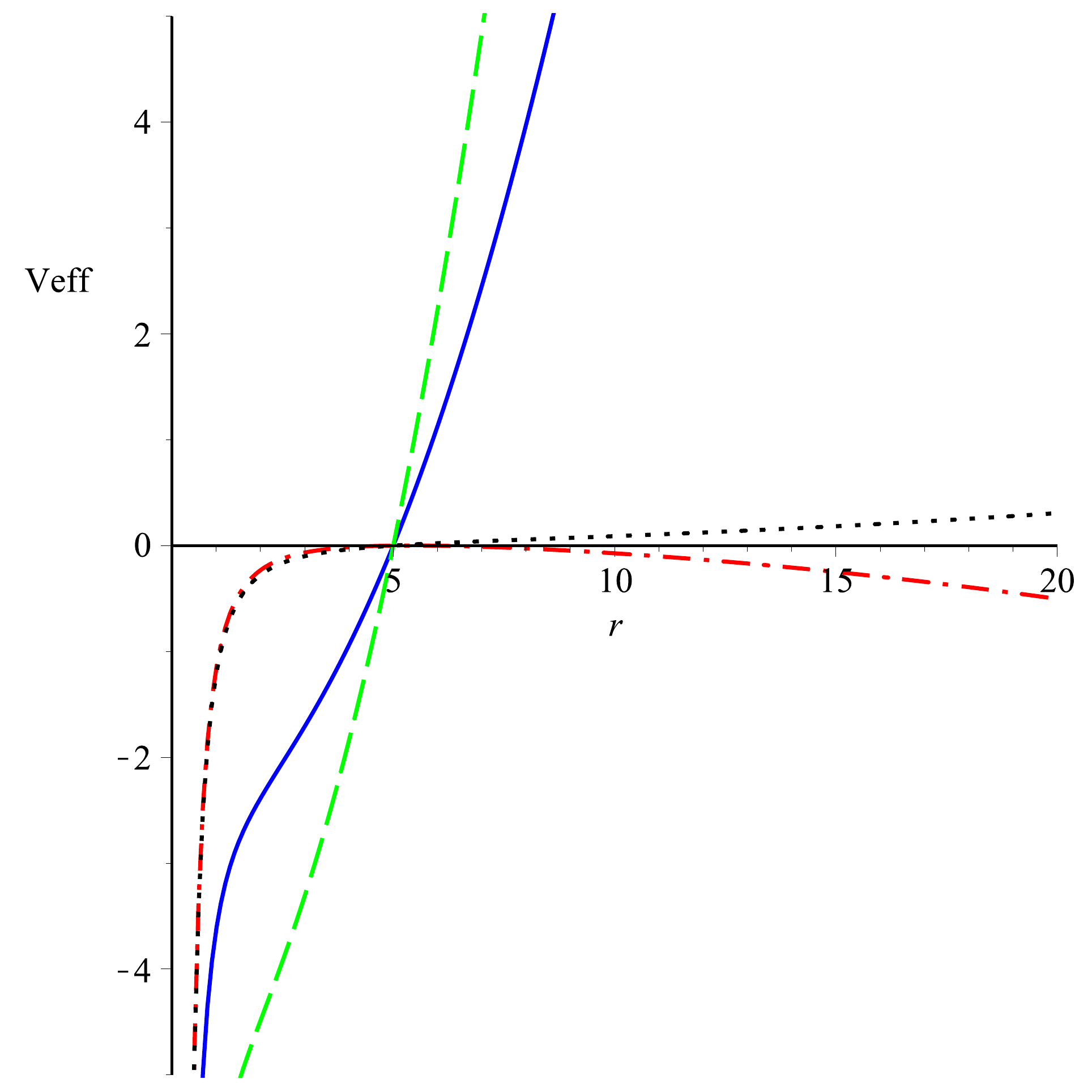}
    \caption{{\scriptsize $m\!=\!1$, $M\!=\!1$, $m_s\!=\!1$, $l\!=\!5$, $\xi\!=\!15$(solid),\\ $\xi\!=\!20$(dash), $\xi\!=\!400$(dashdot), $\xi\!=\!2000$(dot)}}
    \label{fig10}
  \end{subfigure}
  \caption{The effective potential vs $r$ for different values of parameters. In figures a, c, and d the event horizon is located at $r=5$. The plots show the growth of instability by increasing $m, m_{s}$, and $\xi$.}
  \label{f2}
\end{figure}
%@@@@@@@@@@@@@@@
\end{itemize}

%@@@@@@@@@@@@@@@@@@
\subsection{Analytical solution of radial equation}
In order to solve the radial equation (\ref{req}) analytically and determine the QNMs of the scalar field we introduce a dimensionless parameter as
\be \label{zdef} z=1-\frac{r_h^2}{r^2}\,,\qquad 0\le z\le 1\ee
where the equality values $z=0$ and $z=1$ are correspond with the event horizon and asymptotically far region, respectively. The differential equation (\ref{req}) in terms of parameter $z$ becomes
\be \label{zeq} z(1-z) \frac{d^2 R}{dz^2}+(1-z) \frac{dR}{dz}+\left(\frac{A}{z}-\frac{B}{1-z}-C\right)R=0\,,\ee
where the constants are defined as
\be \label{const1} A=\frac{l^2 \omega^2}{4 M}\,,\qquad B=\frac{l^2 m_s^2}{4(l^2-\xi)}\,,\qquad C=\frac{m^2}{4M}\,.\ee
We can rewrite the Eq. (\ref{zeq}) as an Hypergeometric differential equation by using the following redefinition of radial function
\be\label{fred} R(z)=z^{\alpha} (1-z)^\beta F(z)\,,\ee
such that by substituting this in Eq. (\ref{zeq}) the new function $F$ satisfies the following equation
\be\label{heq1} z(1-z) \frac{d^2 F}{dz^2}+\left(1+2\alpha-(1+2\alpha+2\beta)z\right) \frac{dF}{dz}+\left(\frac{A'}{z}-\frac{B'}{1-z}-C'\right)F=0\,,\ee
where now the constants are given in terms of new parameters $\alpha$ and $\beta$ by
\be \label{const2} A'=A+\alpha^2\,,\qquad B'=B+\beta(1-\beta)\,,\qquad C'=C+(\alpha+\beta)^2\,.\ee
To find the solution we must apply some constraints on the coefficient of function $F$ to remove the poles at $z=0$ and $z=1$, so
\be \label{cst} A'=B'=0\,, \ee
where due to these constraints we obtain
\be\label{const3} \alpha=\mp i\frac{l\omega}{2\sqrt{M}}\,,\qquad \beta=\frac12\left( 1\mp\sqrt{1+\frac{l^4 m_s^2}{l^2-\xi}}\right)\,.\ee
Again by definition three new parameters
\be\label{const4} a=\alpha+\beta+i\sqrt{C}\,, \qquad b=\alpha+\beta-i\sqrt{C},,\qquad c=1+2\alpha\,,\ee
which one can easily check they satisfy the relation $c-a-b=1-2\beta$. Gathering the above information the Eq. (\ref{heq1}) finally reduces to a hypergeometric equation \cite{Arfken}
\be\label{heq2} z(1-z) \frac{d^2 F}{dz^2}+\left(c-(1+a+b)z\right) \frac{dF}{dz}+ a b F=0.\ee
The solution of this equation is a linear combination of hypergeometric functions which from the relation (\ref{fred}), the general solution of Eq. (\ref{zeq}) is given by
\be\label{sol1} R=z^{\alpha} (1-z)^{\beta} \left[ C_1\, F(a,b,c;z)+C_2\, F(a-c+1,b-c+1,2-c;z)\right],\ee
where $C_1$ and $C_2$ are constants of integration and $F$s are the hypergeometric functions.
%@@@@@@@@@@@@@@@@@@@@
\subsection{Quasi Normal Modes}
Since we look for the wave functions which are purely ingoing at the horizon $z=0$, the boundary conditions  determine $\alpha=-i\frac{l\omega}{2\sqrt{M}}$.  So, in the near horizon region we have only the contribution of first term in (\ref{sol1}) as
\be R=D z^{\alpha} (1-z)^{\beta} F(a,b,c;z).\ee
The behavior of radial part of the scalar field in the far region, $z=1$, is described by (\ref{fred}) accompanied with the transformation for hypergeometric function given by\cite{AS}
\bea F(a,b,c;z)\!\!\!\!&=&\!\!\!\frac{\Gamma(c) \Gamma(c-a-b)}{\Gamma(c-a)\Gamma(c-b)}\, F(a,b,a+b-c+1;1-z)\nn\\&+&\!\!\!(1-z)^{c-a-b} \,\frac{\Gamma(c) \Gamma(-c+a+b)}{\Gamma(a)\Gamma(b)} F(c-a,c-b,-a-b+c+1;1-z). \eea
From the relations in (\ref{const4}) we have
\bea\label{sol2} R(z) \!\!&\approx&\!\! D\,\frac{(1-z)^{\beta}\Gamma(1+2\alpha) \Gamma(1-2\beta)}{\Gamma(1+\alpha-\beta-i\sqrt{C})\Gamma(1+\alpha-\beta+i\sqrt{C})}\nn\\
&&\!\!\!\!\!+\,D\, \frac{(1-z)^{1-\beta}\Gamma(1+2\alpha) \Gamma(-1+2\beta)}{\Gamma(\alpha+\beta-i\sqrt{C})\Gamma(\alpha+\beta+i\sqrt{C})}.
\eea
Using the relation (\ref{zdef}) we can rewrite the asymptotic  ($r\gg r_{h}$) behavior for radial function as
\be\label{sol3} R(r)\approx Y_+\,\left(\frac{r}{r_h}\right)^{-2\beta}+Y_- \,\left(\frac{r}{r_h}\right)^{-2(1-\beta)}\,,\ee
where the new constants are given by
\bea
 Y_+ \!\!\!\!&\equiv&\!\!\!\! D\,\frac{\Gamma(1+2\alpha) \Gamma(1-2\beta)}{\Gamma(1+\alpha-\beta-i\sqrt{C})\Gamma(1+\alpha-\beta+i\sqrt{C})},\nn\\
 Y_- \!\!\!\!&\equiv&\!\!\!\! D\, \frac{\Gamma(1+2\alpha) \Gamma(-1+2\beta)}{\Gamma(\alpha+\beta-i\sqrt{C})\Gamma(\alpha+\beta+i\sqrt{C})}.
\eea
The first term is an ingoing wave at the asymptotic infinity while the second one represents an outgoing wave.
Since we want the scalar wave function to be purely outgoing, so the first constant $Y_+$ must be zero. From the properties of the Gamma functions in general textbooks of Mathematics (Weierstras's form)\cite{AS}, due to vanishing the first term we obtain
\be \label{qnm1} 1+\alpha-\beta-i\sqrt{C}=-n\,,\qquad or \qquad1+\alpha-\beta+i\sqrt{C}=-n\,,\ee
where $n$ is a non-negative integer number which is called overtone number \cite{Cardoso:2001bb}. Substituting the values of constant parameters from the relations (\ref{const1}) and (\ref{const3}) in (\ref{qnm1}) we achieve the following QNMs for the scalar field
\be\label{qnmhR} \omega_{R}=-\frac{m}{l}-i\frac{2\sqrt{M}}{l}\left(n+\frac12+\frac12\sqrt{\frac{l^2-\xi+m_s^2 l^4 }{(l^2-\xi)}}\right)\,,\ee
\be\label{qnmhL} \omega_{L}=+\frac{m}{l}-i\frac{2\sqrt{M}}{l}\left(n+\frac12+\frac12\sqrt{\frac{l^2-\xi+m_s^2 l^4 }{(l^2-\xi)}}\right)\,.\ee

We have solved Eq. (\ref{seq})  subjected to appropriate boundary conditions (regularity at the horizon and at infinity) by direct integration, looking for eigenvalues $\omega=\omega_{R}+i\,\omega_{I}$. According to the time dependence of ansatz (\ref{ansatz}), stable modes are characterized by $\omega_{I}<0$ and unstable modes by $\omega_{I}>0$. As is obvious from (\ref{qnmhR}) and (\ref{qnmhL}) mostly all the physical parameters have a contribution to the imaginary part of QNMs, thus they play an important role in stability of perturbations around non-rotating BTZ black holes. So as been anticipated increasing the non-minimal coupling constant $\xi$ can destabilizes the black hole. In the table (\ref{tab1},\ref{tab2}) we have given an exact spectrum for the QNM frequencies in terms of different categories of parameters. The frequencies of both tables have negative imaginary parts and they confirm that we have stability and by increasing the scalar mass and coupling constant the absolute value of imaginary parts also increase which state that the system undergoes destabilization, such as observed in the effective potentials.

\begin{table}[H] %\renewcommand{\arraystretch}{2}
\center
\begin{tabular}{|c|c||c|c||c|c||c|c||}\hline
      $n$ & $\omega_{R}$($l=5,m=1$) &$m_{s}$&$\omega_{R}$($l=5,m=1$)&$\xi$&$\omega_{R}$($l=5,m=1$) &$M$&$\omega_{R}$($l=5,m=1$)\\ \hline
0 & 0.7798-0.2000 i &0 &-0.2000-2.400 i&50&0.7798-2.200 i&1&0.3414-2.200 i\\
1&  0.7798-0.6000 i &1&0.3414-2.200 i&150&0.2000-2.200 i&2&0.5656-3.110 i\\
2&  0.7798-1.000 i&2& 0.9372-2.200 i&250&0.0666-2.200 i&3&0.7378-3.810 i\\
3&  0.7798-1.400 i &3&1.520-2.200 i&350&-0.0078-2.200 i&4&0.8828-4.400 i\\
4&  0.7798-1.800 i&4& 2.100-2.200 i&450&-0.0628-2.200 i&5&1.011-4.920 i\\
5&  0.7798-2.200 i &5&2.680-2.200 i&550&-0.1127-2.200 i&6&1.126-5.388 i\\
6&  0.7798-2.600 i &6&3.258-2.200 i&650&-0.2000-2.200 i&7&1.233-5.822 i\\
7&  0.7798-3.000 i &7&3.836-2.200 i&750&-0.2000-2.274 i&8&1.331-6.222 i\\
8&  0.7798-3.400 i&8&4.414-2.200 i &850&-0.2000-2.298 i&9&1.424-6.600 i\\
9&  0.7798-3.800 i&9&4.992-2.200 i &950&-0.2000-2.314 i&10&1.512-6.956 i\\
10&  0.77980-4.200 i&10&5.570-2.200 i&1050&-0.2000-2.324 i&11&1.596-7.298 i \\ \hline
\end{tabular}
\caption[]{$\omega_{R}$ frequencies for different values of parameters in four categories; in the first one we choose $\xi=50,m_{s}=1,M=1$, in the second $n=5,\xi=100,M=1$, in the third $n=5,m_{s}=1,M=1$, and finally in the forth $n=5,\xi=100,m_{s}=1$.}
\label{tab1}
\end{table}

\begin{table}[H] %\renewcommand{\arraystretch}{2}
\center
\begin{tabular}{|c|c||c|c||c|c||c|c||}\hline
      $n$ & $\omega_{L}$($l=5,m=1$) &$m_{s}$&$\omega_{L}$($l=5,m=1$)&$\xi$&$\omega_{L}$($l=5,m=1$) &$M$&$\omega_{L}$($l=5,m=1$)\\ \hline
0 & 1.180-0.2000 i &0 &-0.2000-2.400 i&50&1.180-2.200 i&1&0.7414-2.200 i\\
1&  1.180-0.6000 i &1&0.7414-2.200 i&150&0.6000-2.200 i&2&0.9656-3.110 i\\
2&  1.180-1.000 i&2& 1.337-2.200 i&250&0.4666-2.200 i&3&1.138-3.810 i\\
3&  1.180-1.400 i &3&1.920-2.200 i&350&-0.3922-2.200 i&4&1.283-4.400 i\\
4&  1.180-1.800 i&4& 2.500-2.200 i&450&-0.3372-2.200 i&5&1.410-4.920 i\\
5&  1.180-2.200 i &5&3.080-2.200 i&550&-0.2872-2.200 i&6&1.526-5.388 i\\
6&  1.180-2.600 i &6&3.658-2.200 i&650&-0.2000-2.200 i&7&1.633-5.822 i\\
7&  1.180-3.000 i &7&4.236-2.200 i&750&-0.2000-2.274 i&8&1.731-6.222 i\\
8&  1.180-3.400 i&8&4.814-2.200 i &850&-0.2000-2.298 i&9&1.824-6.600 i\\
9&  1.180-3.800 i&9&5.392-2.200 i &950&-0.2000-2.314 i&10&1.912-6.956 i\\
10&  1.180-4.200 i&10&5.970-2.200 i&1050&-0.2000-2.324 i&11&1.996-7.298 i \\ \hline
\end{tabular}
\caption[]{$\omega_{L}$ frequencies for different values of parameters in four categories;  The parameters are chosen as the values in the caption of table (1).}
\label{tab2}
\end{table}
%@@@@@@@@@@@@@@
\subsection{A comment on the AdS/CFT correspondence}
The AdS/CFT correspondence \cite{Maldacena:1997re,Gubser:1998bc,Witten:1998qj} has led to important pogress in our understanding of the microscopic physics of a class of near extremal black holes. According to this conjecture  for some sectors of three dimensional gravities which are asymptotically AdS$_{3}$, there exists a two dimensional dual conformal field theory\cite{Witten:1998qj}. However, the presence of black holes in AdS spacetime corresponds to turning on a temperature in the dual CFT. The QNMs of black holes correspond to tiny deviations of the thermal equilibrium in dual field theory. Especially for two-dimensional CFT, the left and right sectors are independent. At thermal equilibrium, the two sectors may have different temperatures $(T_{L}, T_{R})$ where in the case of BTZ black holes are given in \cite{Maldacena:1998bw}
\be\label{temps} T_{L}=\frac{r_{+}-r_{-}}{2\pi l}\,,\qquad T_{R}=\frac{r_{+}+r_{-}}{2\pi l}\,.\ee
Consider a small perturbation of an operator with conformal weights $h_{L}$ and $h_{R}$. The system will return to thermal equilibrium exponentially with a characteristic time scale, which is inversely proportional to the imaginary part of the poles of the correlation function of the operator in momentum space \cite{Birmingham:2001hc,Birmingham:2001pj}\cite{Cardy:1986ie,Horowitz:1999jd}, that is, $\tau=\frac{1}{|\omega_{I}|}$. These poles are
\be\label{poles} \omega_{L}=k-4\pi i T_{L} (n+h_{L})\,,\qquad \omega_{R}=-k-4\pi i T_{R} (n+h_{R})\,.\ee

It has been shown in Refs. \cite{Birmingham:2001hc,Birmingham:2001pj,Gubser:1997cm,Cardoso:2001hn} that these poles are in agreement with the QNMs of a minimally coupled scalar field of mass $m_{s}$ around BTZ black holes which the conformal weights of its corresponding operator in the dual CFT are denoted by 
\be\label{cws1} h_L=h_R=\frac12\left(1+\sqrt{1+m_s^2 l^2}\,\right),\ee
 and its conformal dimension and spin are defined by $\Delta\equiv h_{L}+h_{R}=1+\sqrt{1+m_s^2 l^2}$ and $s\equiv h_{L}-h_{R}=0$, respectively\cite{Aharony:1999ti}. Now by comparing the relations in (\ref{poles}) with (\ref{qnmhR}) and (\ref{qnmhL}) we can deduce that there may be some operators in the dual CFT with conformal weights
\be\label{cws2} h_L=h_R=\frac12+\frac12\sqrt{\frac{l^2-\xi+m_s^2 l^4 }{(l^2-\xi)}}\,,\ee
which are in correspondence with the non-minimally coupled massive scalar fields (\ref{ansatz}) in the gravitational sector. As seen, in the limit $\xi\rightarrow 0$ the conformal weights in (\ref{cws2}) lead to the ones in relation (\ref{cws1}) where in this limit, the QNMs of non-minimally coupled scalar field (\ref{qnmhR}) and (\ref{qnmhL}) lead to the QNMs of minimal scalar field of mass $m_{s}$ in Refs. \cite{Birmingham:2001hc,Birmingham:2001pj}. Though we do not consider the details of the dual field theory content here but if we accept the duality be established by virtue of (\ref{poles}) then the limit $\xi\rightarrow 0$ give the same outcome implications in the dictionary. So the imaginary parts of the QNM frequencies in (\ref{qnmhR}) and (\ref{qnmhL}) can determine the decay rate of small perturbations when the scalar field is non-minimally coupled with the Einstein tensor in the vicinity of uncharged non-rotating BTZ black holes in the AdS sector and the return to equilibrium is specified in terms of the left and right timescales $\tau_{L} =\frac{1}{(\omega_{L})_{I}}$ and $\tau_{R} =\frac{1}{(\omega_{R})_{I}}$ in the CFT side.
%@@@@@@@@@@@@@@@@@@@@@@@@@@
%@@@@@@@@@@@@@@@@@@
\section{Greybody factors and Hawking radiation}
\begin{figure}[H]
\centering
\includegraphics[width=5.5cm,height=4cm]{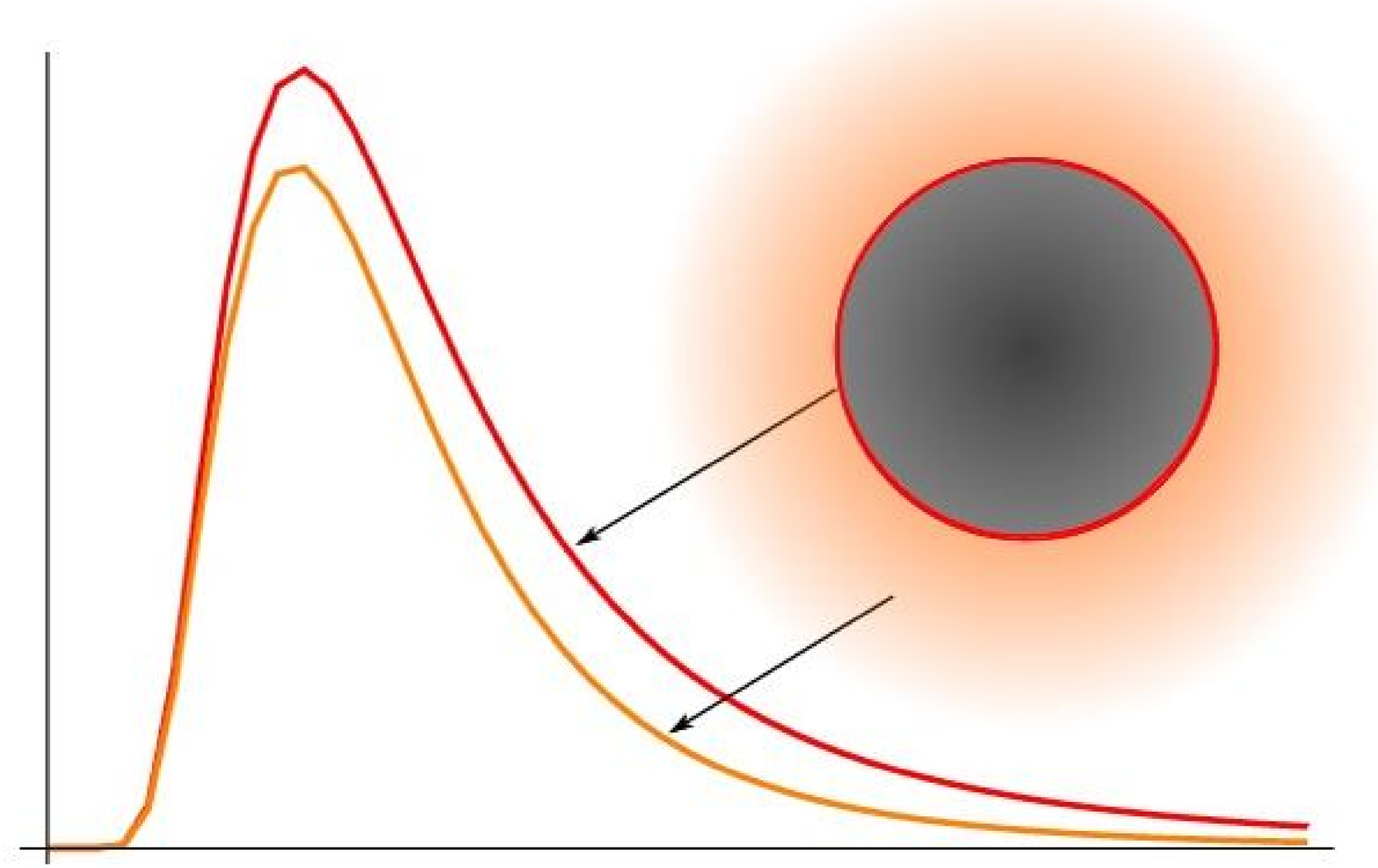}
\caption{The solid red line is for emission rate at the event horizon and the orange one is for geometry around the black hole.}
\label{BH}
\end{figure}

About one half century ago, Hawking showed that black holes can radiate just like a thermal system and due to this property, they produce a connection between the classical gravity and quantum mechanics\cite{Hawking:1974sw,Hawking:1974rv}. This radiation emitted by the black hole is described by a black body spectrum at the event horizon which is consistent with the ingoing boundary condition that we have chosen before. However, far away from the black hole horizon, it will get modified by the black hole geometry. In the picture shown in Fig.(\ref{BH}), we have drown how the geometry around the black hole modifies the hawking radiation at the event horizon. This change in the spectrum is given by the GFs. In a semiclassical approximation these GFs can be calculated by studying the scattering of a field in the black hole background.

The emission rate of a black hole, named as Hawking radiation \cite{Hawking:1974sw}, in the frequency $\omega$ is given by
\be\label{hr1} \Gamma(\omega)=\frac{1}{e^{\beta\omega}\pm1} \frac{d^3 k}{(2\pi)^3},\ee
where $\beta$ is the inverse of the Hawking temperature of the black hole at event horizon and the minus (plus) sign is used when considering bosons (fermions). This formula is valid for both massless and massive particles.

\begin{figure}[H]
\centering
\includegraphics[width=5.5cm,height=4cm]{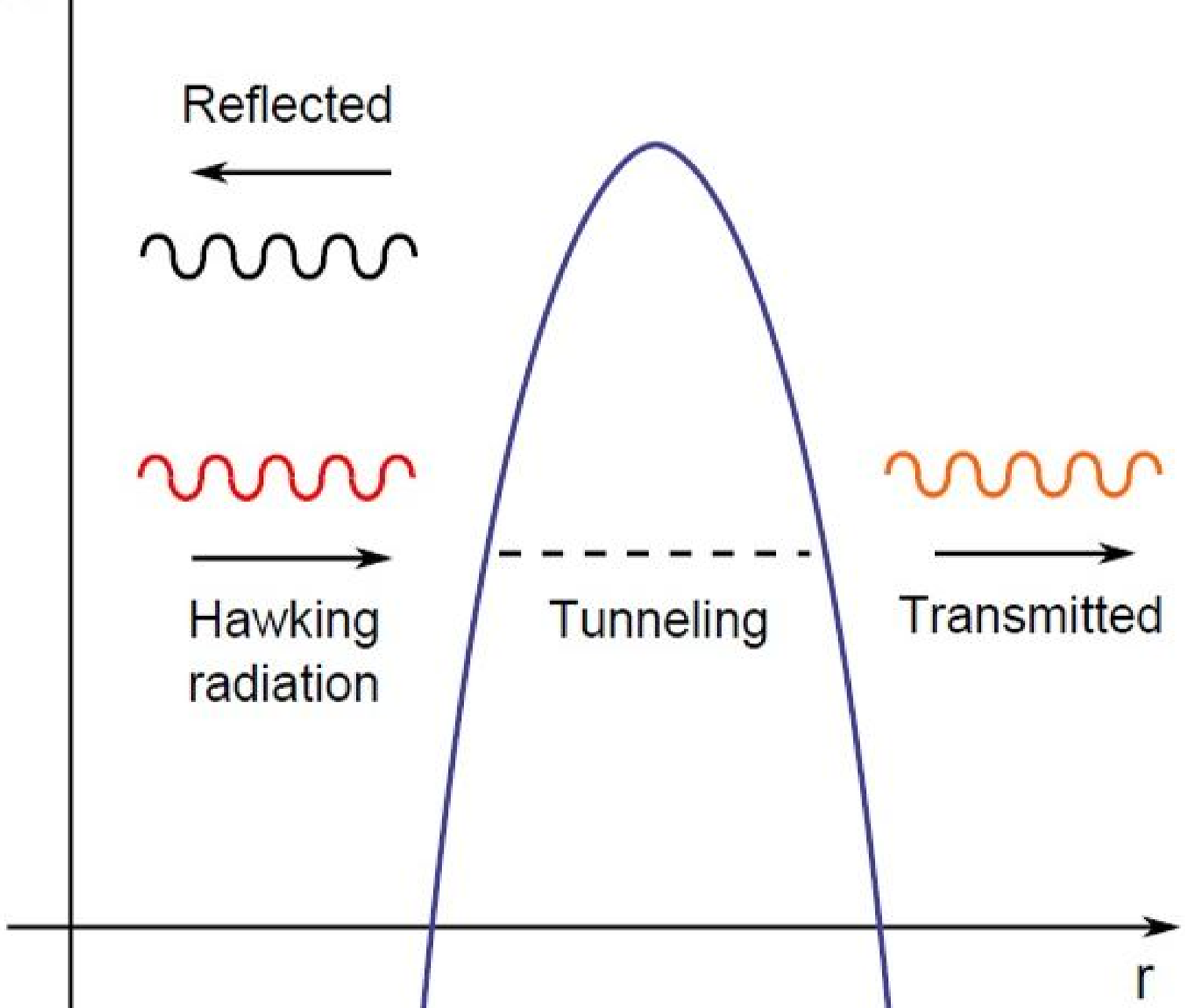}
\caption{Filtering of Hawking radiation by potential barrier caused by black hole geometry.}
\label{HR}
\end{figure}

In general, the geometry of the spacetime surrounding a black hole is non-trivial. Taking this into account, we might imagine that once Hawking radiation is emitted at the event horizon, it will get modified by this non-trivial geometry so that when an observer located very far away from the black hole measures the spectrum, this will no longer be that of a black body. This is indeed the case: the black hole geometry outside the event horizon acts as a potential barrier that filters Hawking radiation, i.e., part of it will be transmitted and will travel freely to infinity, whereas the rest will be reflected back into the black hole. We have depicted this implication in Fig.(\ref{HR}).
We can summarize the previous statements by saying that the spectrum emitted by a black hole that an observer at spatial infinity would measure is given by \cite{Hawking:1974sw}
\be\label{hr2} \Gamma(\omega)=\frac{\gamma_{m}(\omega)}{e^{\beta\omega}\pm1} \frac{d^3 k}{(2\pi)^3},\ee
where $\gamma$ is the so-called greybody factor of angular quantum number $m$, which depends on the frequency of the
particles under consideration and is given by
\be\label{gf1}  \gamma_{m}(\omega)=1-\left|\frac{A^{out}(\omega)}{A^{in}(\omega)}\right|^2\,,\ee
where $A^{in}$ and $A^{out}$ are, respectively, the ingoing and outgoing amplitudes of the wave function and depend on the corresponding frequencies. In general contexts, the left hand side of (\ref{gf1}) is also referred to the absorption cross section, $\mathbb{\sigma}_{abs}(\omega)$, see Refs. \cite{Das:1996we},\cite{Maldacena:1996ix}, \cite{Gubser:1997cm}, and the second term of the right hand side is called the reflection coefficient $\mathcal{R(\omega)}$ \cite{Kanti:2005ja}. So, it is instructive to consider these coefficients and as a consequently the rate of black hole decay (\ref{hr2}), in different regimes of frequencies.

In order to find the GFs we need to consider the behavior of the scalar wave function at the far region. We should estimate the equation at large distances and then match the solution of the differential equation to one we obtained in sec. 3. So, we solve the differential equation (\ref{req}) in the $r\rightarrow \infty$ limit. To this end, it becomes
\be\label{fareq} R''+\frac{3}{r} R'-{\frac {{l}^{4}{m_s}^{2}}{ \left( {l}^{2}-\xi \right) {r}^{2}}}
R=0, \ee
where its solution is given simply by
\be \label{farsol}  R(r)\sim X_1 r^{-1-\sqrt{1+\frac{l^4\,{m_s}^2}{l^2-\xi}}}+X_2 r^{-1+\sqrt{1+\frac{l^4\,{m_s}^2}{l^2-\xi}}}.\ee

Here we have used the approximation mark $\sim$ instead of equality to show the result  is an asymptotic solution. The two relations (\ref{sol3}) and (\ref{farsol}) for $R(r)$ in the large $r$ limit are called ``stretched'' solutions. Using the minus sign for $\beta$ in relation (\ref{const3}) similar to one chosen in the previous section, it was shown that both of these relations have the same power-law behavior and their smooth matching is straightforward. So, identifying the coefficients of the same powers of $r$, we arrive at equalities
\bea \label{farcoeff}  X_1=Y_+\!\!\!&\equiv&\!\!\!D\,\frac{\Gamma(1+2\alpha) \Gamma(1-2\beta)}{\Gamma(1+\alpha-\beta-i\sqrt{C})\Gamma(1+\alpha-\beta+i\sqrt{C})},\nn\\
X_2= Y_-\!\!\! &\equiv&\!\!\!D\, \frac{\Gamma(1+2\alpha) \Gamma(-1+2\beta)}{\Gamma(\alpha+\beta-i\sqrt{C})\Gamma(\alpha+\beta+i\sqrt{C})}.
\eea
Hence, the GFs and the reflection coefficient from the relation (\ref{gf1}) are
\be\label{gf2}  \gamma_{m}(\omega)=1-\mathcal{R}\,,\qquad \mathcal{R}=\left|\frac{X_2}{X_1}\right|^2,\ee
furthermore, by substituting the parameters from (\ref{const1}) and  (\ref{const3}) in the constant coefficients (\ref{farcoeff}) we obtain
\be \label{refc} \mathcal{R}=\frac{\cosh\left[\frac{\pi}{2}\left(\frac{l \omega}{\sqrt{M}}-\sqrt{\frac{l^4 {m_s}^2}{\xi-l^2}-1}+\frac{m}{\sqrt{M}}\right)\right]}{\cosh\left[\frac{\pi}{2}\left(\frac{l \omega}{\sqrt{M}}+\sqrt{\frac{l^4 {m_s}^2}{\xi-l^2}-1}+\frac{m}{\sqrt{M}}\right)\right]}\,\frac{\cosh\left[\frac{\pi}{2}\left(\frac{l \omega}{\sqrt{M}}-\sqrt{\frac{l^4 {m_s}^2}{\xi-l^2}-1}-\frac{m}{\sqrt{M}}\right)\right]}{\cosh\left[\frac{\pi}{2}\left(\frac{l \omega}{\sqrt{M}}+\sqrt{\frac{l^4 {m_s}^2}{\xi-l^2}-1}-\frac{m}{\sqrt{M}}\right)\right]}\,,\ee
in which we have used the following relations for the gamma functions \cite{Arfken},
\be |\Gamma(iz)|^2=\frac{\pi}{z\sinh(\pi z)}\,,\qquad \left|\Gamma\left(\frac12+iz\right)\right|^2=\frac{\pi}{cosh(\pi z)}. \ee
Notice that according to the conditions (\ref{excs}) for the parameter $\beta$ we have
\be 1+\frac{l^4 m_s^2}{l^2-\xi}<0\,,\ee
where we have used this condition to achieve the relation (\ref{refc}).
It has been shown in \cite{Das:1996we} that the GFs of arbitrary quantum number in the case of asymptotically flat black holes in d dimensions vanishes in the zero frequency limit. The validity of this result in d=4 dimensions for the minimal and non-minimal coupling of a scalar field with Ricciscalar are given in \cite{Brady:1996za} and \cite{Chen:2010ru}.

In the low energy limit we can expand the GFs as
\be \label{gf3} \gamma=\gamma_{0}+\gamma_1\,\omega+\mathcal{O}(\omega^2)\,, \ee
where
\bea \label{gf4}
\gamma_0&\!\!\!=\!\!\!&0,\nn\\
\gamma_1&\!\!\!=\!\!\!&\frac{\pi l}{\sqrt{M}}\frac{2\sinh{\left[\frac{\pi}{2}\sqrt{\frac{l^4 {m_s}^2}{\xi-l^2}-1}\,\right]} \cosh{\left[\frac{\pi}{2}\frac{m}{\sqrt{M}}\right]}}{\cosh^2{\left[\frac{\pi}{2}\sqrt{\frac{l^4 {m_s}^2}{\xi-l^2}-1}\,\right]} \, \cosh^2{\left[\frac{\pi}{2}\frac{m}{\sqrt{M}}\right]}-\sinh^2{\left[\frac{\pi}{2}\sqrt{\frac{l^4 {m_s}^2}{\xi-l^2}-1}\,\right]}\, \sinh^2{\left[\frac{\pi}{2}\frac{m}{\sqrt{M}}\right]}}\nn\\
&\!\!\!=\!\!\!&\frac{\pi l}{\sqrt{M}}\left[\tanh\left(\frac{\pi}{2}\frac{m}{\sqrt{M}}+\frac{\pi}{2}\sqrt{\frac{l^4 {m_s}^2}{\xi-l^2}-1}\right)-\tanh\left(\frac{\pi}{2}\frac{m}{\sqrt{M}}-\frac{\pi}{2}\sqrt{\frac{l^4 {m_s}^2}{\xi-l^2}-1}\right)\right]\,.
\eea

\begin{figure}[H]
\centering
  \begin{subfigure}{0.4\textwidth}
    \includegraphics[width=\textwidth]{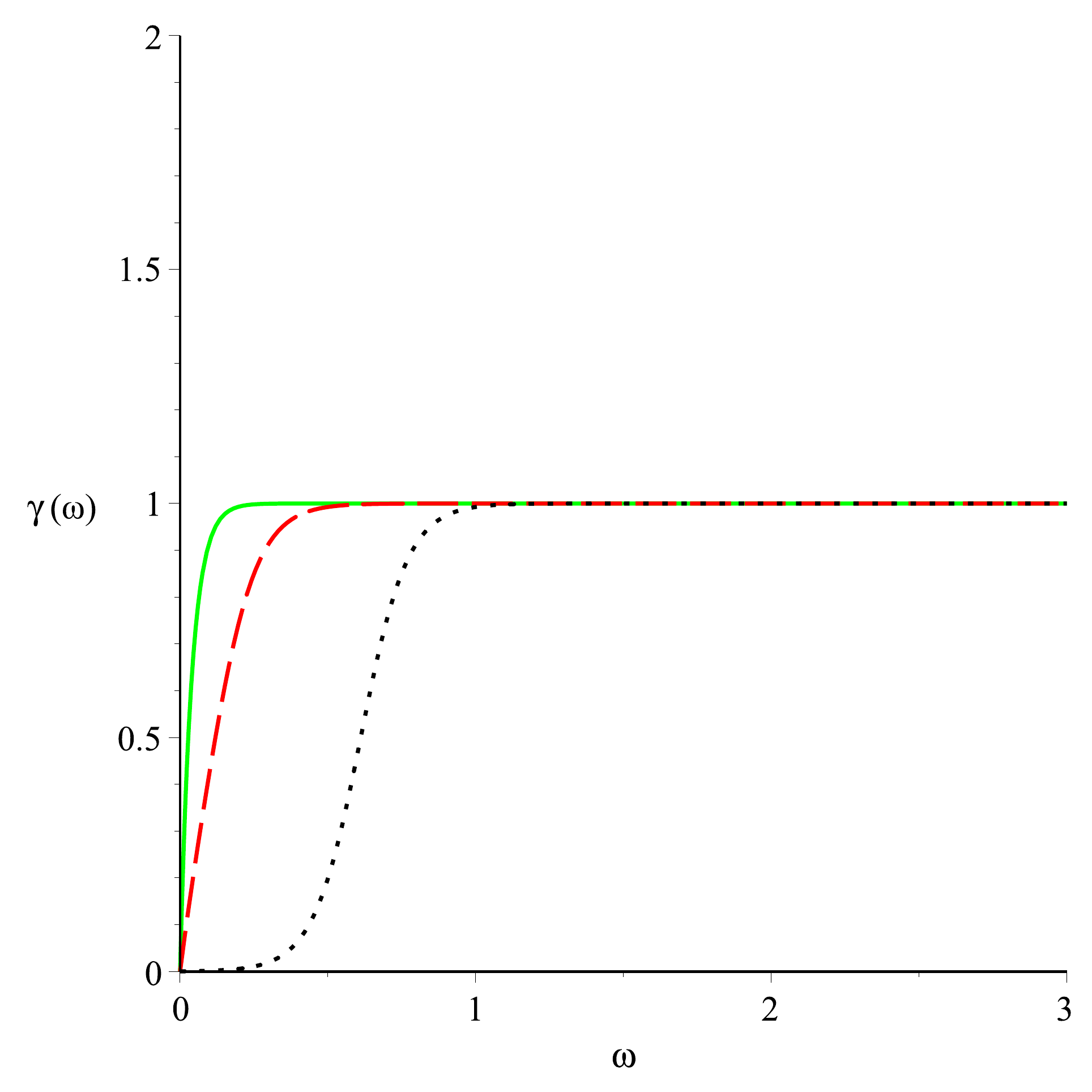}
    \caption{{\scriptsize $l\!=\!4$, $M\!=\!1$, $\xi\!=\!50$, $l\!=\!5$, $m_s\!=\!1$, \\ $m\!=\!0$(solid), $m\!=\!3$(dash), $m\!=\!5$(dot)}}
    \label{fig11}
  \end{subfigure}
  \hspace{5mm}
  \begin{subfigure}{0.4\textwidth}
    \includegraphics[width=\textwidth]{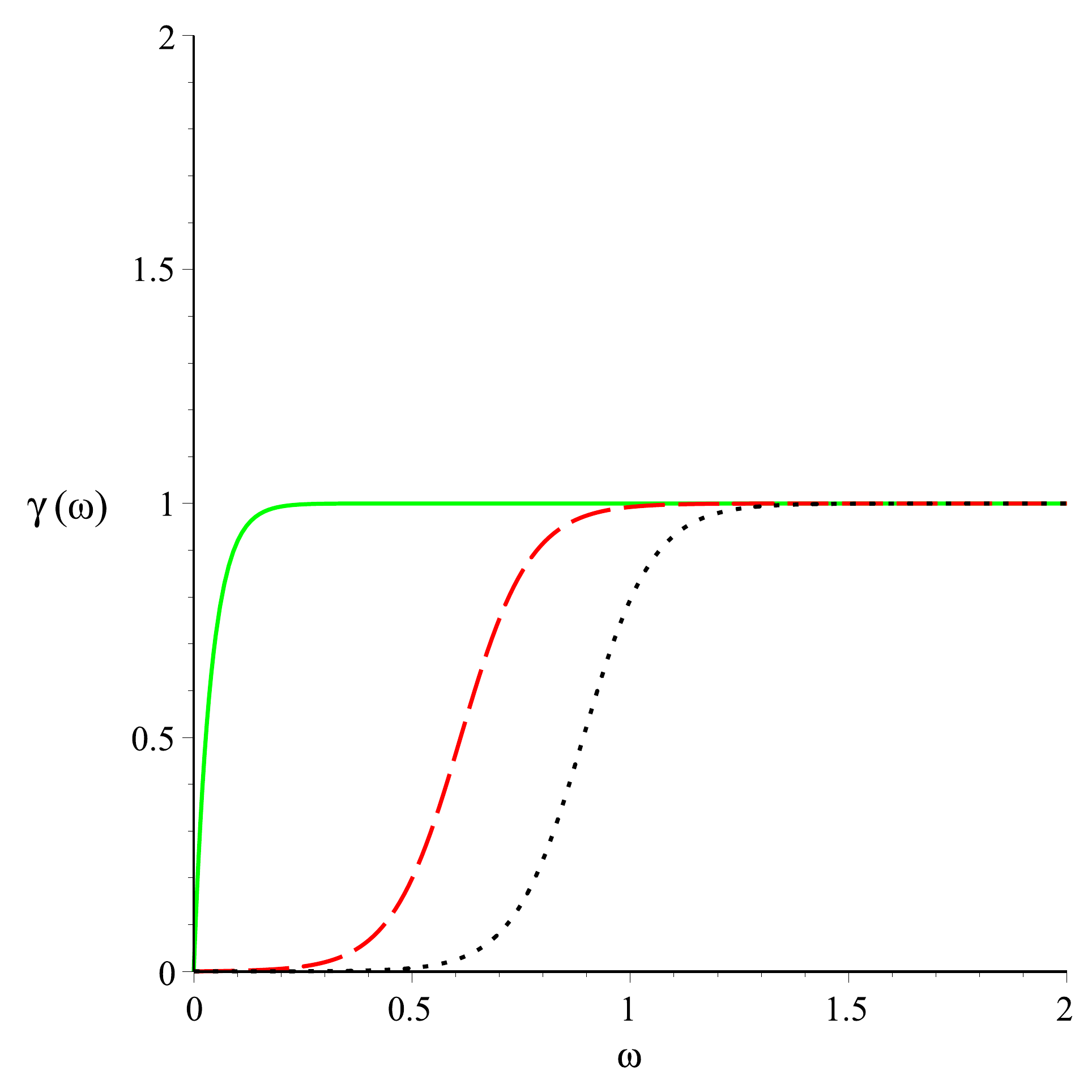}
    \caption{{\scriptsize $m\!=\!5$, $l\!=\!4$, $M\!=\!1$, $m_{s}\!=\!1$,\\ $\xi\!=\!20$(solid), $\xi\!=\!50$(dash), $\xi\!=\!100$(dot)}}
    \label{fig12}
  \end{subfigure}
    \caption{The greybody factors for variable coupling $\xi$ and quantum number $m$ vs. $\omega$ }
    \label{f3}
\end{figure}

As seen, in the low energy limit, the GFs depend linearly on the frequency $\omega$ according to (\ref{gf4}). Of course, if we choose the constant parameters such that $\xi=l^2 (m_{s}^2 l^2+1)$, then this linear dependence also vanishes. Similar to previous considerations for the effective potential, we now are able to study these factors numerically by plotting them from the relations (\ref{gf2}) and (\ref{refc}) in terms of particle and spacetime properties. For instance, in Fig.(\ref{f3}) we have sketched two diagrams to show the dependence of GFs to different values of coupling constant $\xi$, in Fig.(\ref{fig11}), and angular quantum number $m$,  in Fig.(\ref{fig12}). It is evident  in both diagrams that as we increase $\xi$ and $m$, the GFs were suppressed at low frequencies.

\begin{figure}[H]
\centering
  \begin{subfigure}{0.4\textwidth}
    \includegraphics[width=\textwidth]{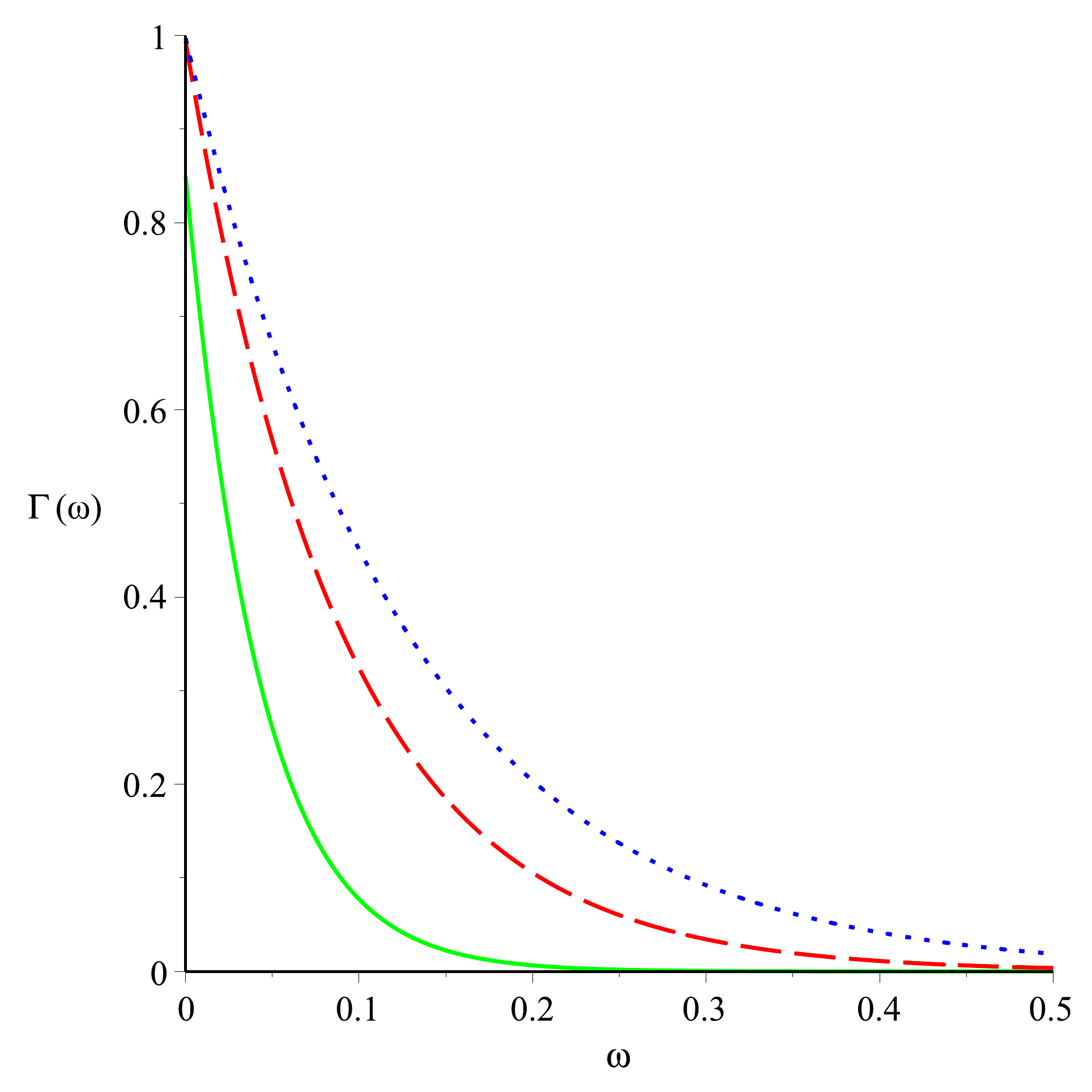}
    \caption{{\scriptsize $m\!=\!2$, $l\!=\!4$, $\xi\!=\!50$, $m_s\!=\!2$,\\ $M\!=\!1$(solid), $M\!=\!5$(dash), $M\!=\!10$(dot)}}
    \label{fig13}
  \end{subfigure}
  \hspace{5mm}
  \begin{subfigure}{0.4\textwidth}
    \includegraphics[width=\textwidth]{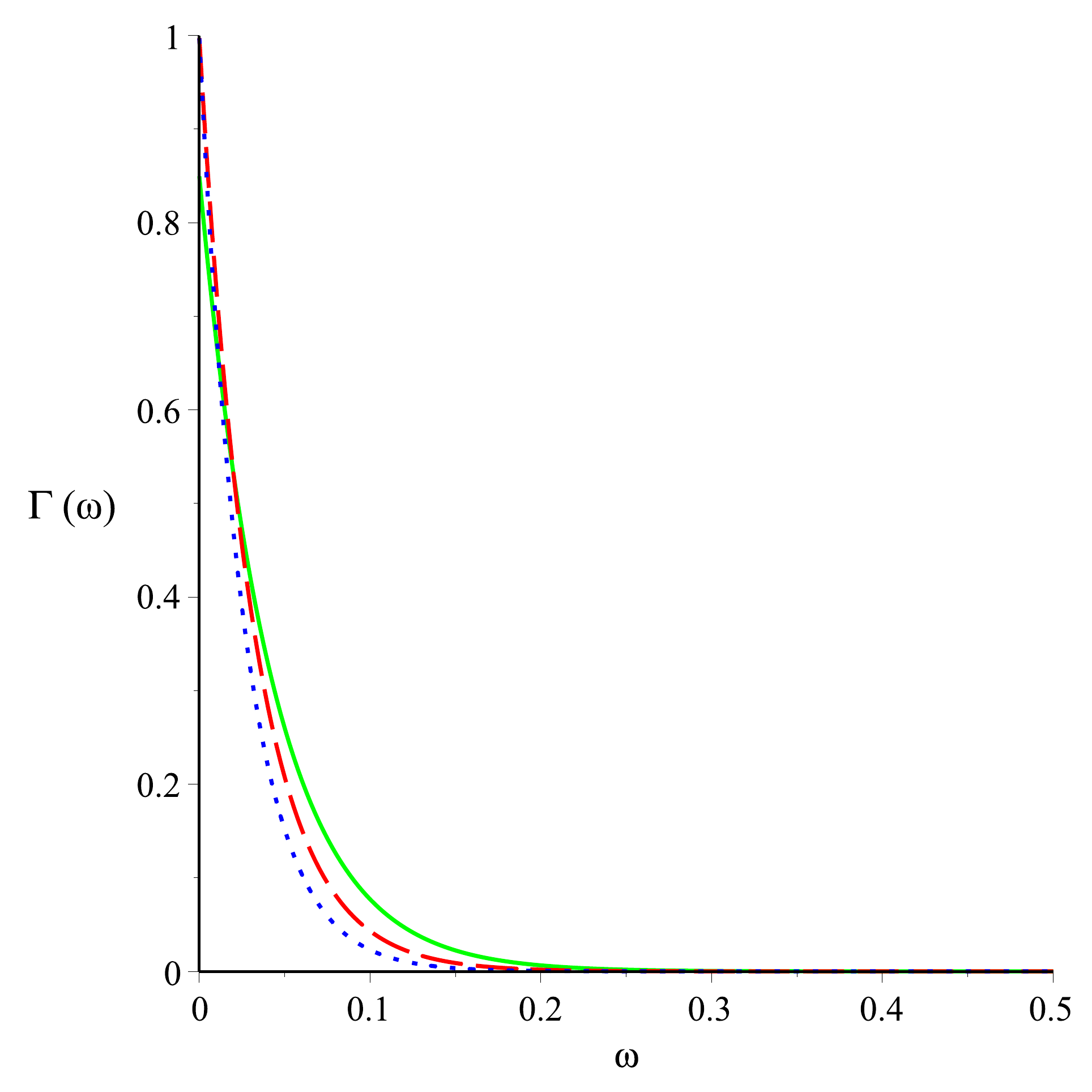}
    \caption{{\scriptsize $m\!=\!2$, $M\!=\!1$, $\xi\!=\!50$, $m_s\!=\!1$,\\ $l\!=\!4$(solid), $l\!=\!5$(dash), $l\!=\!6$(dot)}}
    \label{fig14}
  \end{subfigure}
  \caption{The Hawking radiation for BTZ black hole of different $M$ and $l$ vs $\omega$}
  \label{f4}
\end{figure}

In Fig.(\ref{f4}) we study the emission rate of Hawking radiation for the black hole properties as a function of frequency $\omega$. We plot the relation (\ref{gf2}) by using the Hawking temperature (\ref{HT}) computed for the BTZ black hole only for variable black hole mass $M$ and AdS radius of geometry around the black hole $l$. As stated in sec. 3, the latter is related to the cosmological constant term in asymptotically AdS like black holes. By comparing the plots in Figs. (\ref{fig13}) and (\ref{fig14}), we observed that their behaviors are in contrast to each other, that is, by increasing the values of $l$ the rate of emission is suppressed while by increasing the mass of black hole it is amplified. We have plotted the behavior for other parameters and observed that the diagrams have similar suppression manner to Fig. (\ref{fig14}), but we do not bring them here.

%%%%%%%%%%%%%%%%%%%%%%%%%%%%%%%%%%%%%%%%%%%%
\section{Conclusion}
In this work we have studied the impressive role of a black hole geometry on the propagation of a massive scalar field when it emitted in the gravitational background of a non-rotating BTZ black hole in three-dimensional spacetime. We considered a theory in which a scalar field coupled to Einstein tensor of gravitational background non-minimally. We have shown that the scalar field equation in its radial part changed to a Schr$\ddot o$dinger-like equation with an effective potential by using a tortoise change of coordinate $r$. We have plotted the behavior of this potential for different values of particle and spacetime properties in Figs. (\ref{f1}) and (\ref{f2}). It has shown that there is a constraint on the parameters for which we have a local maximum for the potential as a necessary point to consider the greybody factors. According to the Fig.~(\ref{f1}), when we increased the coupling constant regarding to (\ref{excs}), the potential shows different behavior and the local maximum disappears by increasing the cosmological constant. We showed in Fig. (\ref{f2}) that the BTZ black hole is stable, but by increasing either the mass of scalar field $m_{s}$ or coupling constant $\xi$, it destabilizes.

We obtained an exact analytical spectrum for the QNMs of massive scalar field perturbations around the BTZ black hole by solving the radial equation of motion in terms of hypergeometric functions and then the results were given in the Tables (\ref{tab1},\ref{tab2}). The results have shown that the imaginary parts in all ranges of parameters are negative and consequently the system is stable, but by increasing the constants of the scalar field and the background we encountered destabilization. 
We have asserted that if the AdS/CFT correspondence be established by virtue of the relations in (\ref{poles}), which have been demonstrated in \cite{Gubser:1997cm}, then by using the QNM frequencies in (\ref{qnmhR}) and (\ref{qnmhL}) we provided expressions for the conformal weights of dual operators for corresponding scalar fields in the CFT side given by (\ref{cws2}). 

 We have also studied the Hawking radiation rates of BTZ black holes and showed that these emission rates were modified by some greybody factors which are defined because of the geometry around the black hole. We derived the GFs in the low energy limit and it was shown that to zeroth order of $\omega$ they vanished, while to first order we have a non-vanishing term as a function of scalar field and black hole parameters. We investigated this order can be zero as well, for some choice of coupling constant. However, we have depicted these factors from (\ref{gf2}) for scalar characters in Fig. (\ref{f3}). Though, we could plot this factor for different parameters but we only sketched it in terms of $\xi$ and $m$ and the plots showed a suppression by increasing their values. In contrast, we plotted the Hawking radiation in terms of black hole parameters $M$ and $l$ in Fig. (\ref{f4}) which showed opposite behavior by increasing their values.

This calculations have been done for an Einsteinian metric which its Einstein tensor is proportional to the metric, so the equation of motion can be transform to a Klein-Gordon field equation with an effective mass. But we considered the non-minimal coupling constant separately in our work which is useful to compare with non-Einsteinian metrics such as the warped AdS black hole or Lifshits black holes in three-dimensional spacetime as future works.

%%%%%%%%%%%%%%%%%%%%%%%%%%%%%%%%%%%%%%%%%%%%%%%
%\appendix
%%%%%%%%%%%%%%%%%%%%%%%%%%%%%%%%%
%\section*{Acknowledgment}
%@@@@@@@@@@@@@@@@@@@@@@@@@@@@@@@@@@@@@@@@@@@@@@@@@@@@@@@@@@@@@@@@@@@@@@@@@@@@@@@@@@@@
%%%%%%%%%%%%%%%%%%%%%%%%%%%%%

\end{document}